\documentclass[12pt,reqno]{amsart}
\allowdisplaybreaks[4]
\usepackage{graphicx} 
\usepackage{amssymb}
\usepackage{amsmath}
\usepackage{cite}
\usepackage[hang,nooneline]{subfigure}                        
\newcounter{fig}   

\begin{document}
\title[Rogue wave ]
{Rogue waves in nonlinear Schr\"odinger models with variable coefficients:
application to Bose-Einstein condensates}
\author{J. S. He, ${ }^{1}$ E. G. Charalampidis, ${ }^{2,3,4}$ P. G. Kevrekidis, ${ }^{4}$ D. J. Frantzeskakis ${ }^{5}$
 }
\dedicatory { ${}^{1}$ \ Department of Mathematics, Ningbo University,
Ningbo , Zhejiang 315211, P.\ R.\ China \\
              ${}^{2}$ \ School of Civil Engineering, Faculty of Engineering,\newline
Aristotle University of Thessaloniki, Thessaloniki 54124, Greece \\
              ${}^{3}$ \ Institut f\"ur Physik, Universit\"at Oldenburg, Postfach 2503, D-26111 Oldenburg, Germany \\
              ${}^{4}$ \ Department of Mathematics and Statistics, University of Massachusetts,  Amherst, \newline Massachusetts 01003-4515, USA \\
              ${}^{5}$ \ Department of Physics, University of Athens, Panepistimiopolis, Zografos, Athens 15784, Greece
              }

\begin{abstract}

We explore the form of rogue
wave solutions in a select set of case examples
of nonlinear Schr\"odinger equations with variable coefficients. We focus
on systems with constant dispersion, and present three different models that
describe
atomic Bose-Einstein condensates in different experimentally relevant
settings. For these
models, we identify exact rogue wave solutions. Our analytical findings are
corroborated by
direct numerical integration of the original equations, performed by two
different schemes.
Very good agreement between numerical results and analytical predictions for the emergence of the rogue waves is identified. Additionally, the nontrivial
fate of small numerically induced perturbations to the exact rogue wave
solutions is also discussed.

\end{abstract}

\maketitle



\section{Introduction}
Extreme wave events, initially reported in ocean seafarer stories and later observed
by satellite surveillances \cite{k1}, have become a subject of increased
interest in physical oceanography \cite{k2}. In this context, pertinent water waves are known as
rogue waves (alias ``freak'' or ``extreme'' waves), and are tentatively defined as
waves whose height is more than twice the significant wave height, which is itself defined
as the averaged of the highest third of the waves in a time series \cite{k2}; therefore, rogue waves
are large waves for a given sea state. The conditions that cause rogue waves to grow enormously
in size are not fully understood up to now, although there exists an ongoing effort
(see, e.g., Refs.~\cite{effort,akhmediev01}) and discussions \cite{discussion} regarding this problem.

Obviously, the elucidation of the mechanisms underlying the formation and dynamics of rogue waves
is a subject of fundamental scientific interest; as such, it has attracted significant attention
not only in the more standard ocean-surface-dynamical problem \cite{k2}, but also in other physical
contexts. Indeed, there exists a vast amount of theoretical work in various fields ranging
from optics (see, e.g., the recent work \cite{optth,xuhe1,xuhe2,lihe2,hefokas1} and references therein) and atomic
Bose-Einstein condensates (BECs) \cite{vvk}, to plasmas \cite{pl}, laser-plasma
interactions \cite{gv}, atmospheric dynamics \cite{atm}, and even econophysics \cite{econ}; see also the recent short review of~\cite{solli2,onorato}.

In addition, many experimental observations of rogue waves have been reported in different settings,
including nonlinear optics \cite{opt1,opt2,opt3}, mode-locked lasers \cite{laser},
superfluid helium \cite{He}, hydrodynamics \cite{hydro}, Faraday surface ripples
\cite{fsr}, parametrically driven capillary waves \cite{cap}, and plasmas \cite{plasma}.

Interestingly enough, in many of the above contexts, the models that are used to
describe rogue waves are variants of a universal nonlinear evolution equation, namely
the nonlinear Schr\"odinger (NLS) equation. The latter is ubiquitous in wave packet
propagation in nonlinear dispersive media, describing, e.g., the amplitude of deep water
waves, the electric field envelope in optical media, the macroscopic wave function in
superfluids, etc. Importantly, rogue wave events may be triggered by instabilities (e.g., the
modulational instability \cite{MI}) that are present in NLS models. On the other hand,
the NLS equation itself supports solutions that reproduce the
qualitative characteristics of rogue waves to a highly satisfactory extent: indeed,
earlier pioneering works by Peregrine \cite{pere}, Kuznetsov \cite{kuz}, Ma \cite{ma},
and Akhmediev \cite{akh} (and also the work of Dysthe and Trulsen \cite{dt}), succeeded in
constructing rogue-wave-like rational solutions of the NLS model. Importantly, the relevance
of this analytical toolbox for rogue waves, was later established in various experiments
carried out in different physical contexts (see, e.g., Refs.~\cite{opt2,hydro,plasma}).

In this work, we study rogue wave solutions in NLS models with variable coefficients.
Such models have gained attention, mainly due to their relevance in different physical contexts
and, especially, in
 atomic Bose-Einstein condensates (BECs) \cite{BEC,emergent}
and nonlinear optics \cite{kiag}.
In particular, in the physics of BECs, an NLS [alias Gross-Pitaevskii (GP)] equation
usually incorporates an external spatially-dependent potential, which can also become
time-dependent \cite{BEC}; additionally, the nonlinearity coefficient (which is proportional to
the $s$-wave scattering length) may vary in time or/and in space by employing
external magnetic \cite{mFRM} or optical \cite{oFRM} fields close to Feshbach resonances.
On the other hand, in the context of optics, NLS equations with varying dispersion have been
studied in the context of dispersion managed systems \cite{tur}; furthermore, in nonlinear
optical systems, the transverse (to the propagation direction) refractive index profile appears
as a spatially-dependent potential term --similar to that in BECs-- while modulation of the
nonlinearity coefficient is possible too (see, e.g., the review \cite{boris} and references therein).
Localized solutions and solitons of NLS models with variable coefficients have been studied
in many works, while connection of such models with integrable ones has been investigated
as well (see, e.g., \cite{serkin1,jvv1,jvv2,laks,yan1,lihe1,xuhe3,hefokas2} and references therein). Notice that
``nonautonomous rogons'', namely rogue waves in a variable coefficient NLS model, were also
studied recently \cite{nonaut}.

Here, we consider a fairly general $(1+1)$-dimensional NLS model,
incorporating an external potential (thus resembling the GP equation in BECs); the latter, along
with the dispersion and nonlinearity coefficients are assumed to be functions of both independent
variables. Then, following the analysis of Ref.~\cite{lihe1}, we obtain
exact rogue wave solutions of the considered NLS model by
using a general transformation for the unknown field, which involves a ``seed
solution'' that satisfies the usual NLS model, expressed in appropriate new variables.
The latter, along with auxiliary amplitude and phase functions involved in the transformation
of the unknown field, depend on a set of auxiliary functions and parameters. Focusing on the case of
constant dispersion, we show that proper choices of these auxiliary functions and parameters,
lead to three specific models, which can be used to describe atomic BECs in different, yet
experimentally tractable, settings.
For these models, which have not been presented or studied before
(to the best of our knowledge) in the context of rogue waves,
exact analytical rogue wave solutions are presented. Our analytical results are corroborated by
direct numerical simulations, that are performed in the original NLS model, and have been based
on two different integration schemes. For each of the three models, we find
very good agreement
between the two schemes, as well as with the analytically predicted form of each rogue wave solution. However, these computations raise a number of nontrivial
concerns about the robustness of these solutions. In particular, we will
see that in most of the considered cases, for times beyond the formation
of the Peregrine soliton (which will represent a rogue wave in
our case examples), spontaneously a modulational instability of the
background emerges that produces an expanding sequence of bright
solitary waves in the system. We will thus also briefly discuss
the relevant observations in light of the particular form of
our numerical computations (and the examination of two distinct
high order numerical schemes).

The paper is structured as follows. In Section 2 we introduce our model, the analytical methodology,
as well as the rogue wave solutions; we also present and discuss the above mentioned three
specific models that find application in the physics of BECs, as well as their rogue wave solutions.
Section~3 is devoted to a detailed numerical study of these three models, and to a comparison
between our numerical methods and analytical predictions. Finally, Section~4 presents our
conclusions.

\section{Analytical considerations}

\subsection{Model and outline of the analytical method}

We start from
a generic $(1+1)$-dimensional NLS equation with variable coefficients,
which is expressed in dimensionless form as follows:
\begin{equation}
\label{eqvcnlse}
i\frac{\partial \psi}{\partial t}+\frac{D}{2} \frac{\partial^2 \psi}{\partial x^2}
-g|\psi|^2\psi-V\psi=0,
\end{equation}
where $\psi(x,t)$ is a complex field, which represents the macroscopic wave function in BECs
or the electric field envelope in optics; note that in the latter context, $t$ denotes
propagation distance, and $x$ represents either retarded time (for pulses in optical fibers)
or transverse spatial coordinate (for beams in waveguides) \cite{kiag}.
Furthermore, $D$ and $g$ denote the dispersion and nonlinearity coefficients,
while $V$ is an external potential; the latter denotes the trap confining the atoms
in BECs \cite{BEC}, or the linear part of the transverse refractive index profile in optics
\cite{kiag}. Here, we will consider the case $D=D(x,t)$, $g=g(x,t)$ and $V=V(x,t)$, i.e., the
dispersion and nonlinearity coefficients, as well as the external potential, are real
functions of $x$ and $t$.

Variants of Eq.~(\ref{eqvcnlse}) have been studied in Refs.~\cite{serkin1,jvv1,jvv2},
in the contexts of quasi one-dimensional (1D) BECs and nonlinear optical systems.
Particularly, in the case of BECs, the external trap $V$ is naturally spatially inhomogeneous
and can also become time-dependent by using time-varying magnetic or optical fields \cite{BEC},
while the inhomogeneity in $g$ (i.e., in the inter-atomic collision
dynamics) can be induced by means of the Feshbach-resonance management technique; this relies on a
direct control of the $s$-wave scattering length in BECs by employing external (temporally and/or
spatially) varying magnetic \cite{mFRM} or optical \cite{oFRM} fields close to Feshbach resonances.
On the other hand, in the case of optical systems, $V$ and $g$ (i.e., the linear and nonlinear
parts of the refractive index) may be varying in artificial setups, such
as the light-induced photonic lattices \cite{segev} or the structure
composed by layers of silica alternating with empty gaps \cite{martin}. Furthermore,
the dispersion coefficient $D$ may also vary, e.g., in the case of dispersion-managed optical
fiber systems and lasers \cite{tur}.

Following the analysis of Ref.~\cite{lihe1}, we introduce the transformation:

\begin{equation}
\label{mapforvcnlse}
\psi(x,t)=q(X,T)p(x,t)\exp[i\phi(x,t)],
\end{equation}
and reduce Eq.~(\ref{eqvcnlse}) to the usual nonlinear
Schr\"odinger equation satisfied by the auxiliary field $q=q(X,T)$:
\begin{equation}
\label{standardnls}
i\frac{\partial q}{\partial T}+\frac{1}{2}\frac{\partial^2
q}{\partial X^2}+s |q|^2q=0,
\end{equation}
where $s=\pm 1$. The new variable $X=X(x,t)$,
along with the field $p=p(x,t)$ and the phase $\phi=\phi(x,t)$,
as well as $D$, $g$ and $V$, are determined explicitly by means
of a set of arbitrary functions $f_1(t)$, $f_2(t)$, $f_3(t)$, $T(t)$, and
$F(x)$, with a simple condition $F(x)f_1(t)>0$ (see~\cite{lihe1}
and also our specific examples below).
By choosing the above arbitrary functions,
the dispersion and nonlinearity coefficients $D$ and $g$, as well as the potential $V$ in Eq.~(\ref{eqvcnlse}), can be ``designed''
and analytically determined.
Therefore, solutions of Eq.~(\ref{eqvcnlse}) can be found by means of solutions of the
standard NLS of Eq.~(\ref{standardnls}). Here, as we will show below, using the transformation of Eq.~(\ref{mapforvcnlse}) and a rogue wave solution of Eq.~(\ref{standardnls}), we can
determine rogue wave solutions of the original model of
Eq.~(\ref{eqvcnlse}).

\subsection{Rogue waves for systems with constant dispersion}

We now focus on NLS systems with constant dispersion $D$, which are relevant to the
context of atomic BECs. Based on the results of Ref.~\cite{lihe1}, $D={\rm const.}$
can be achieved by choosing $F(x)=c_1$. In such a case, the new variables $X(x,t)$ and $T(t)$,
the field $p(x,t)$ and phase $\phi(x,t)$, as well as the dispersion, nonlinearity and potential functions $D$, $g$ and $V$ of Eq.~(\ref{eqvcnlse}), are given in terms of three real arbitrary
functions, $f_1(t)$, $f_2(t)$ and $f_3(t)$, and three real constants, $c_1$, $c_2$ and $c_3$
(under the condition $c_1f_1(t)>0$), as follows:
\begin{align}
&X(x,t)= c_1 f_1(t)x+f_3(t),
\quad
T(t)=c_2\int f_1^2(t)dt+c_3 ,
\label{eq1parametersforvcnlse}
\\
&p(x,t)=\sqrt{\frac{f_1(t)}{c_1}},
\quad
\phi(x,t)=-\dfrac{c_1}{2c_2} \frac{c_1 \dot{f}_1(t)x^2+2\dot{f}_3(t)x }{f_1(t)}
+f_2(t),
\label{eq3parametersforvcnlse}
\\
&D=\dfrac{c_2}{c_1^2}={\rm const.},
\quad
g=-sc_2c_1 f_1(t),
\quad
V(x,t)= v_2 x^2+v_1 x+v_0,
\end{align}
where dots denote derivatives with respect to $t$, and the coefficients $v_{k}$ $(k=0,1,2)$ appearing 
in the expression of the potential $V(x,t)$ are given by:
\begin{eqnarray}
v_2&=&-\frac{c_1^2}{2c_2}\frac{2\dot{f}_1^2(t)-f_1(t)\ddot{f}_1(t)}{f_1^2(t)},
\label{v2} \\
v_1&=&-\frac{c_1}{c_2}\frac{2\dot{f}_1(t)\dot{f}_3(t)-f_1(t)\ddot{f}_3(t)}{f_1^2(t)},
\label{v1} \\
v_0&=&-\frac{1}{2c_2}\frac{2c_2 f_1^2(t)\dot{f}_2(t)+\dot{f}_3^2(t) }{f_1^2(t)}.
\label{v0}
\end{eqnarray}


We now assume that the seed solution $q(X,T)$ of the transformation (\ref{mapforvcnlse}) is
a rogue wave solution of the NLS Eq.~(\ref{standardnls}); here, we
consider the Peregrine soliton, which has the form (see, e.g., Ref.~\cite{akhmediev01}):
\begin{equation}\label{rwofnlse}
q(X,T) =\left[ 1 - \frac{{4(1 + 2iT)}}{{1 + 4{T^2} + 4{X^2}}}\right]\exp(iT).
\end{equation}
Then, the respective solution $\psi(x,t)$ of Eq.~(\ref{eqvcnlse}) is given by:
\begin{eqnarray}
\label{newslofvcnlse}
\psi(x,t)&=&\sqrt{\frac{f_1(t)}{c_1}}
\left[ 1- \frac{4\left(1+ 2i (c_{2}\int f_1^2(t)dt +c_3) \right)}
{1+4\left(c_2 \int f_1^2(t)dt + c_3\right)^2+4\left(c_1 f_1(t)x +f_3(t) \right)^2}
\right] \nonumber \\
&\times&
\exp\left[i \left(c_2 \int f_1^2(t)dt - \frac{c_1}{2c_2}
\frac{c_{1}\dot{f}_1(t)x^2+2\dot{f}_3(t)x}{f_1(t)}
+f_2(t)+ c_3 \right) \right].
%
\end{eqnarray}
It is clear that if $f_1(t)$ [and, thus, the field $p(x,t)$] is a bounded function of $t$, then
Eq.~(\ref{newslofvcnlse}) describes a rogue wave solution of Eq.~(\ref{eqvcnlse}). Notice that
still other rogue wave solutions can be found, upon choosing the seed solution $q$ to be, e.g.,
the second-order rogue wave of Ref.~\cite{akhmediev01},
the Kuznetsov-Ma soliton \cite{kuz,ma}, the Akhmediev breather \cite{akh,dt}, and so on.


Notice that our approach for deriving rogue wave solutions of Eq.~(\ref{eqvcnlse}) is, arguably,
simpler as compared, e.g., with the methodology of Ref.~\cite{serkin1}, that involves the solution
of a complicated Riccati equation for the coefficients of the NLS, which is not solvable in general.

\subsection{Specific models and their rogue wave solutions}

Having described our analytical approach, we now proceed by presenting three specific case examples,
namely three particular versions of Eq.~(\ref{eqvcnlse}) and their rogue wave solutions. Here, we will
present the analytical results for these cases, while the next section will be devoted to the corresponding
direct numerical simulations.

\subsubsection{Rogue wave on a periodic background.}


Our first case example corresponds to the choice:
$f_1(t)=2+\sin(t)$, $f_2(t)=f_3(t)=1$, and $c_1=c_2=c_3=1$. In this case, for $s=+1$ in
Eq.~(\ref{standardnls}), we find that the field $p$ is given by $p=\sqrt{2+\sin(t)}$,
while the coefficients of Eq.~(\ref{eqvcnlse}) are: $D=1$, $g= -(2 + \sin(t))$, and
$V=-[\big(2 \cos^2(t)  + 2 \sin(t) +\sin^2(t) \big) x^2][2(2 + \sin(t))^2]^{-1}$.
In other words, Eq.~(\ref{eqvcnlse}) takes the form:
%
\begin{eqnarray}
&i\dfrac{\partial \psi}{\partial t}+\dfrac{1}{2}\dfrac{\partial
^2\psi}{\partial x^2}+ [2 +\sin(t)]|\psi|^2\psi = \dfrac{1}{2}\Omega^2(t)x^2 \psi,
\label{model1} \\
&\Omega^2(t)= -\dfrac{2 \cos^2(t)  + 2 \sin(t) +\sin^2(t)  }{(2 + \sin(t))^2}.
\label{om1}
\end{eqnarray}
Physically speaking, this equation models a BEC, with a scattering length (i.e., nonlinearity
coefficient) being periodically modulated in time; this can be achieved by using a periodic external
magnetic or optical field near a Feshbach resonance \cite{mFRM,oFRM}, as mentioned above.
Notice that since the nonlinearity coefficient is always positive in
the context of Eq.~(\ref{model1}), the interatomic interactions
are always attractive (as in the case of $^7$Li \cite{Lisol1,Lisol2} or $^{85}$Rb \cite{Rbsol}
atoms). Furthermore, the external potential $V$ is characterized by a normalized frequency
$\Omega$ which is also time-periodic, and can either be confining,
when the relevant prefactor  $\Omega^2(t)>0$, or
expulsive, for $\Omega^2(t)<0$ (note that such an expulsive potential was used in the experiment of
Ref.~\cite{Lisol2}).

\begin{figure}[tbp]
      \includegraphics[height=0.28\textwidth]{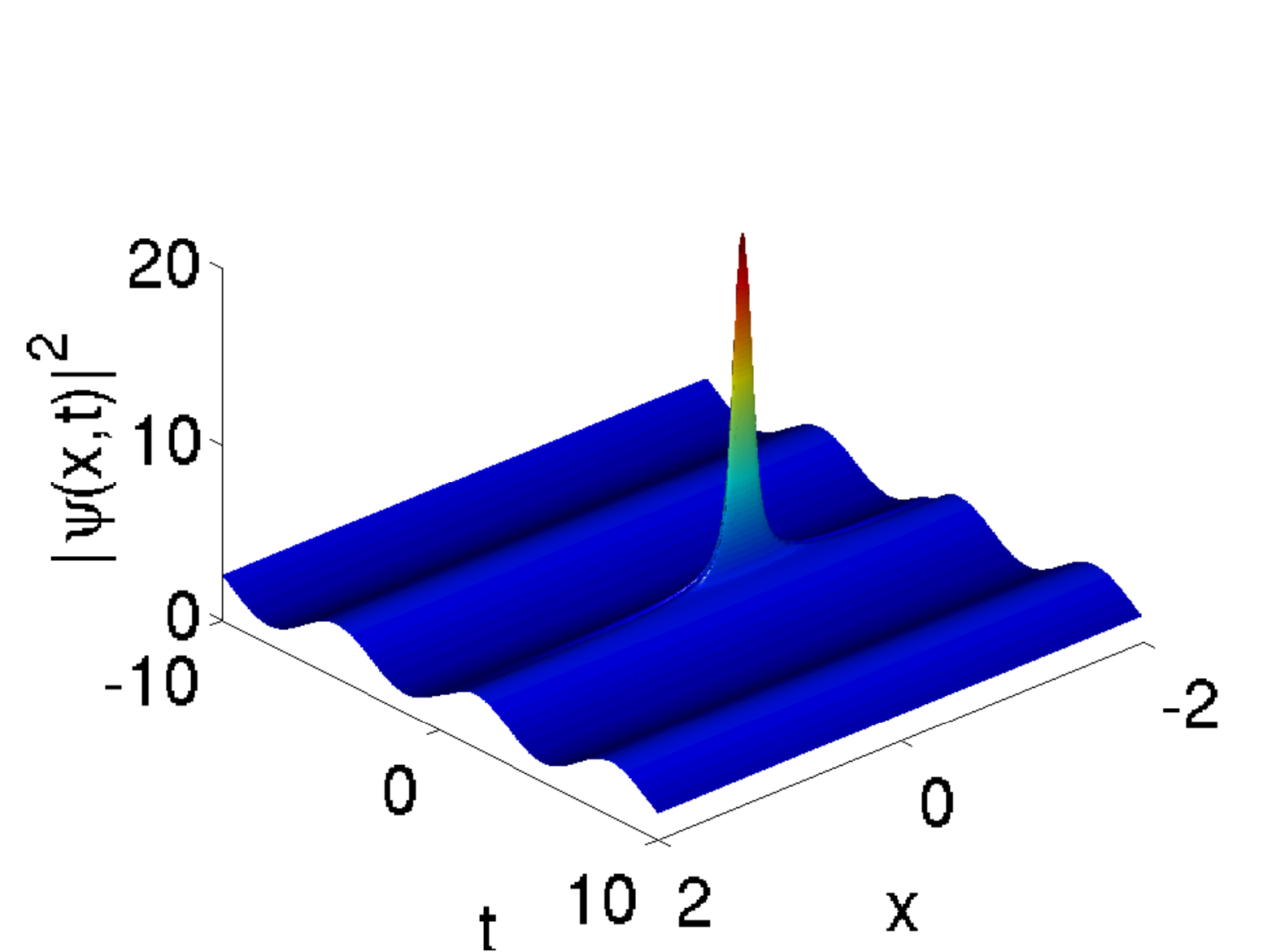}
      \includegraphics[height=0.28\textwidth]{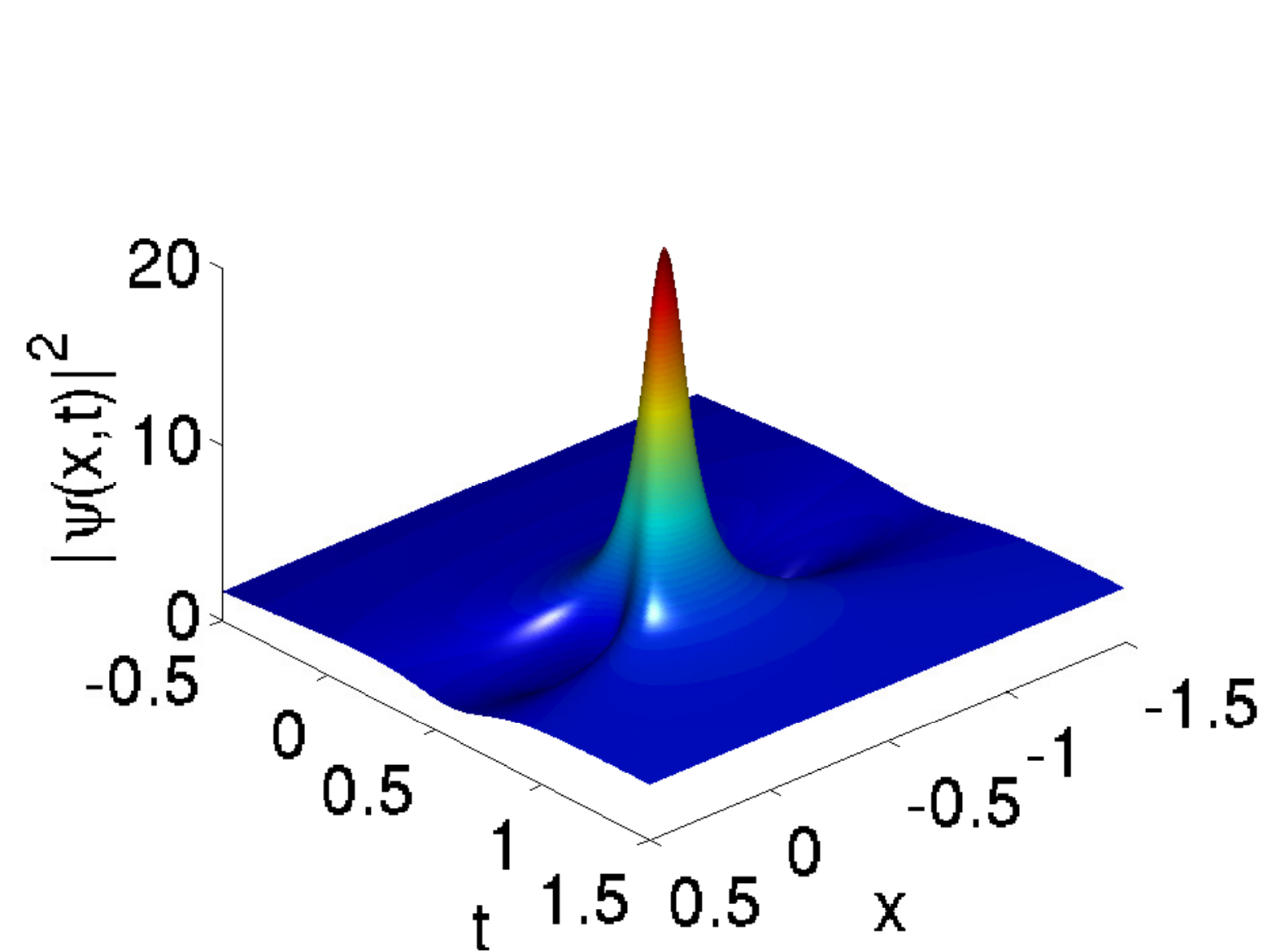}
\caption{The density profile of the rogue wave with a periodic background.
The left panel shows
both the background and the rogue wave, while the right panel shows in more detail the local profile
of the rogue wave.}
      \label{Ffig1}
 \end{figure}

In the case under consideration, it is straightforward to find the specific form of the
rogue wave solution, using Eq.~(\ref{newslofvcnlse}). Since its functional form is rather
complicated, we opt not to show it here. Instead, it is quite useful to illustrate the
density (square modulus) $|\psi|^2$ of this solution, which is shown in Fig.~\ref{Ffig1}.
It is clearly observed that the rogue wave exists on a periodic background (left panel),
and is characterized by a sharp peak followed by density depressions (right panel).

%


\subsubsection{Rogue wave on a monotonically decreasing background.}

We now consider the following choice of the arbitrary functions and parameters:
$f_1(t)= \exp(-t)$, $f_2(t)= f_3(t)=0$, and $c_1=c_3=1$, $c_2=2$. Then, for a focusing
nonlinearity ($s=+1$) in Eq.~(\ref{standardnls}), we find that the field $p$ is given by
$p=\exp(-t/2)$, while the coefficients of Eq.~(\ref{eqvcnlse}) are:
$D=2$, $g=-2\exp(-t)$, and $V=-x^2/4$. Thus, Eq.~(\ref{eqvcnlse}) now becomes:
\begin{equation}
\label{model2}
i\dfrac{\partial \psi}{\partial t}+\dfrac{\partial ^2\psi}{\partial x^2}
+2e^{-t}|\psi|^2\psi +\dfrac{1}{4}x^2\psi=0.
\end{equation}
The above model may describe the dynamics of an attractive BEC (composed by
$^7$Li \cite{Lisol1,Lisol2} or $^{85}$Rb \cite{Rbsol} atoms), under the action
of a purely expulsive potential, described by the last term in the left-hand side of
Eq.~(\ref{model2}). The scattering length (i.e., the nonlinearity coefficient) is
again time-dependent --and this can be achieved by time-varying magnetic \cite{mFRM} or optical
\cite{oFRM} fields close to Feshbach resonance. In this case, the magnitude of the scattering
length assumes an exponentially decreasing in time form, which can be realized by a ramp down
of the applied external fields (such a ramp down of magnetic fields was used in the experiments
of Refs.~\cite{Lisol1,Lisol2}).

\begin{figure}[tbp]
      \includegraphics[height=0.28\textwidth]{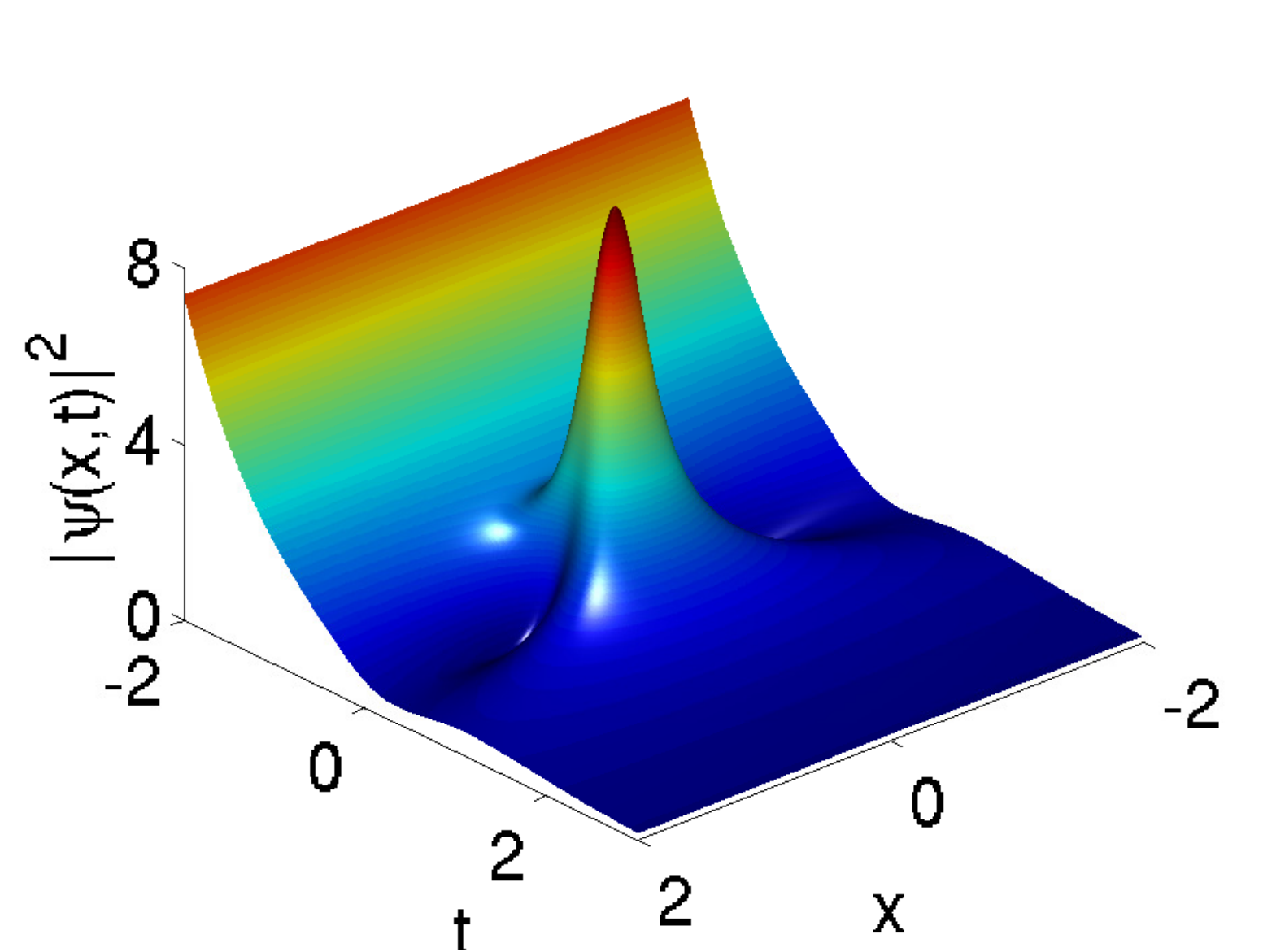}
       \includegraphics[height=0.28\textwidth]{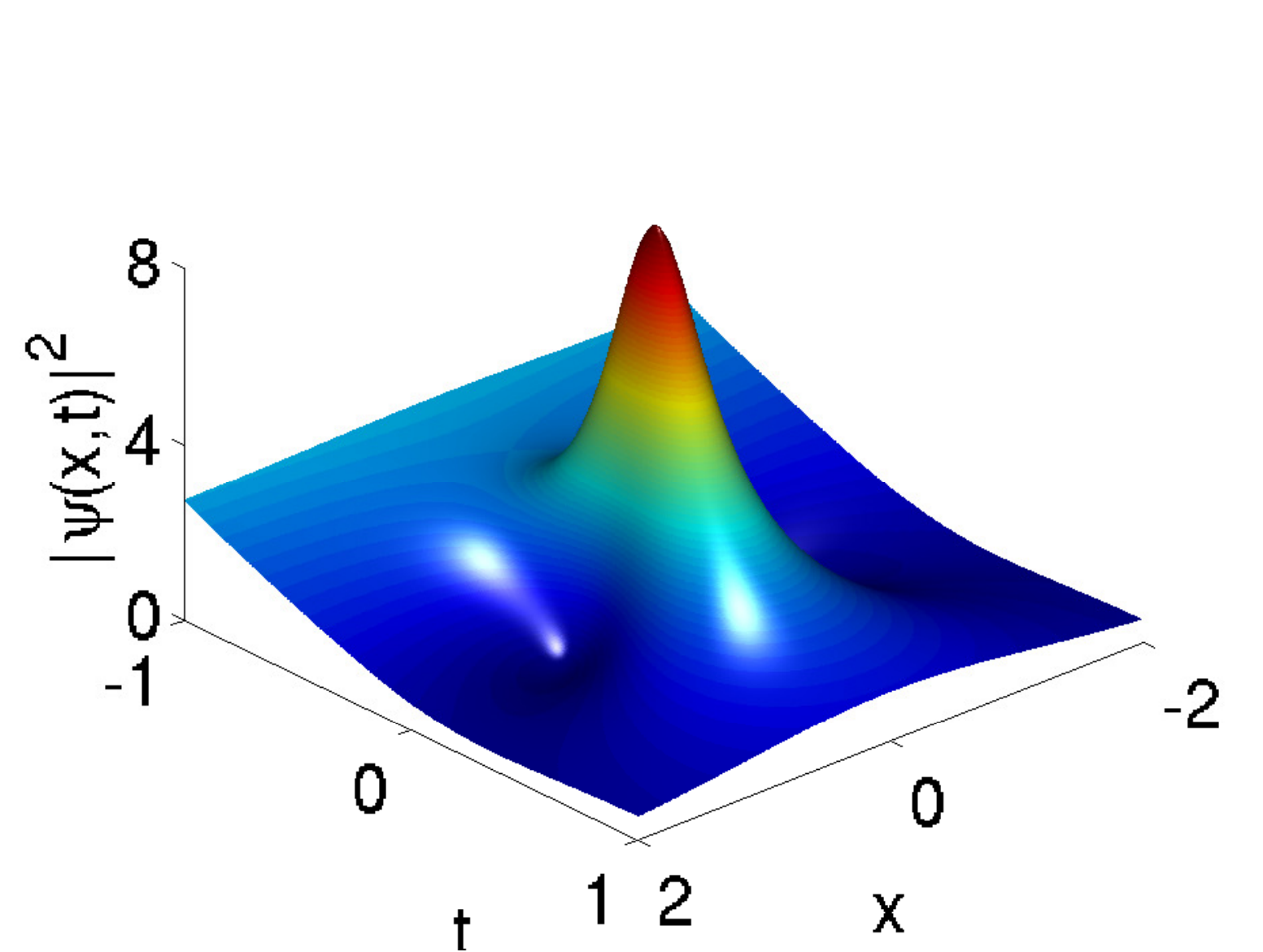}
\caption{The density profile of the rogue wave with a  monotonically decreasing background. The
left panel shows both the background and the rogue wave, while the right panel shows in more detail the local profile of the rogue wave.}
\label{Ffig2}
\end{figure}

As before, the profile of the rogue wave solution for the above mentioned choice
can be found analytically by substituting the relevant functions and parameters in
Eq.~(\ref{newslofvcnlse}). In this case, it is easy to find the density $|\psi|^2$ of the rogue
wave, which is given by:
\begin{equation}\label{solutionofmodel2}
|\psi {|^2}= \dfrac{e^{-t}(65e^{8t}
+8e^{6t}x^2+16e^{4t}x^4+32x^2e^{2t}-64e^{4t}x^2-144e^{6t}+136e^{4t}+16-64e^{2t})}
{(5e^{4t}+4x^2e^{2t}+4-8e^{2t})^2}.
\end{equation}
It is clear that this rogue wave exists on top of a monotonically decreasing background,
as imposed by the form of the field $p$ (recall that $p=\exp(-t/2)$ in this case).
The density profile of the rogue wave on top of this background, along with a
focused snapshot showing its local profile in more detail, are respectively shown in the left and
right panels of Fig.~\ref{Ffig2}.

\subsubsection{Twisted rogue wave on a constant background.}
%
Our last case example, corresponds to the following choice:
$f_1(t)= 1$, $f_2(t)=-(1/12)[3\cos(t)\sin(t)+3t]$, $f_3(t)=-\sin(t)$, and
$c_1=c_2=1$, $c_3=0$; then, for a focusing nonlinearity in Eq.~(\ref{standardnls}) (i.e., for
$s=+1$), we find that the field $p$ is constant, $p=1$, while the dispersion, nonlinearity and
potential in Eq.~(\ref{eqvcnlse}) are respectively given by:
$D=1$, $g=-1$, and $V=x\sin(t)$. Thus, Eq.~(\ref{eqvcnlse}) becomes:
%
\begin{equation}
\label{model3}
i\dfrac{\partial \psi}{\partial t}+\dfrac{1}{2}\dfrac{\partial
^2\psi}{\partial x^2}+|\psi|^2\psi -x\sin(t) \psi=0.
\end{equation}
The above equation is actually a Gross-Pitaevskii model, describing the dynamics of a
BEC with attractive interactions (e.g., a $^7$Li \cite{Lisol1,Lisol2} or a $^{85}$Rb
\cite{Rbsol} BEC), which evolves in a linear (in space) potential, whose amplitude
is sinusoidally modulated in time. In earlier BEC experiments, such a linear potential
was actually realized by a gravitational one \cite{kas}, while in more recent experiments
with BECs loaded in optical lattices, linear potentials periodically modulated in time were also
implemented by using laser beams \cite{ol}. It should also be noted that theoretical studies on
non-autonomous BEC solitons in such time-dependent linear potentials have been reported
as well \cite{fx}.

\begin{figure}[tbp]
      \includegraphics[height=0.28\textwidth]{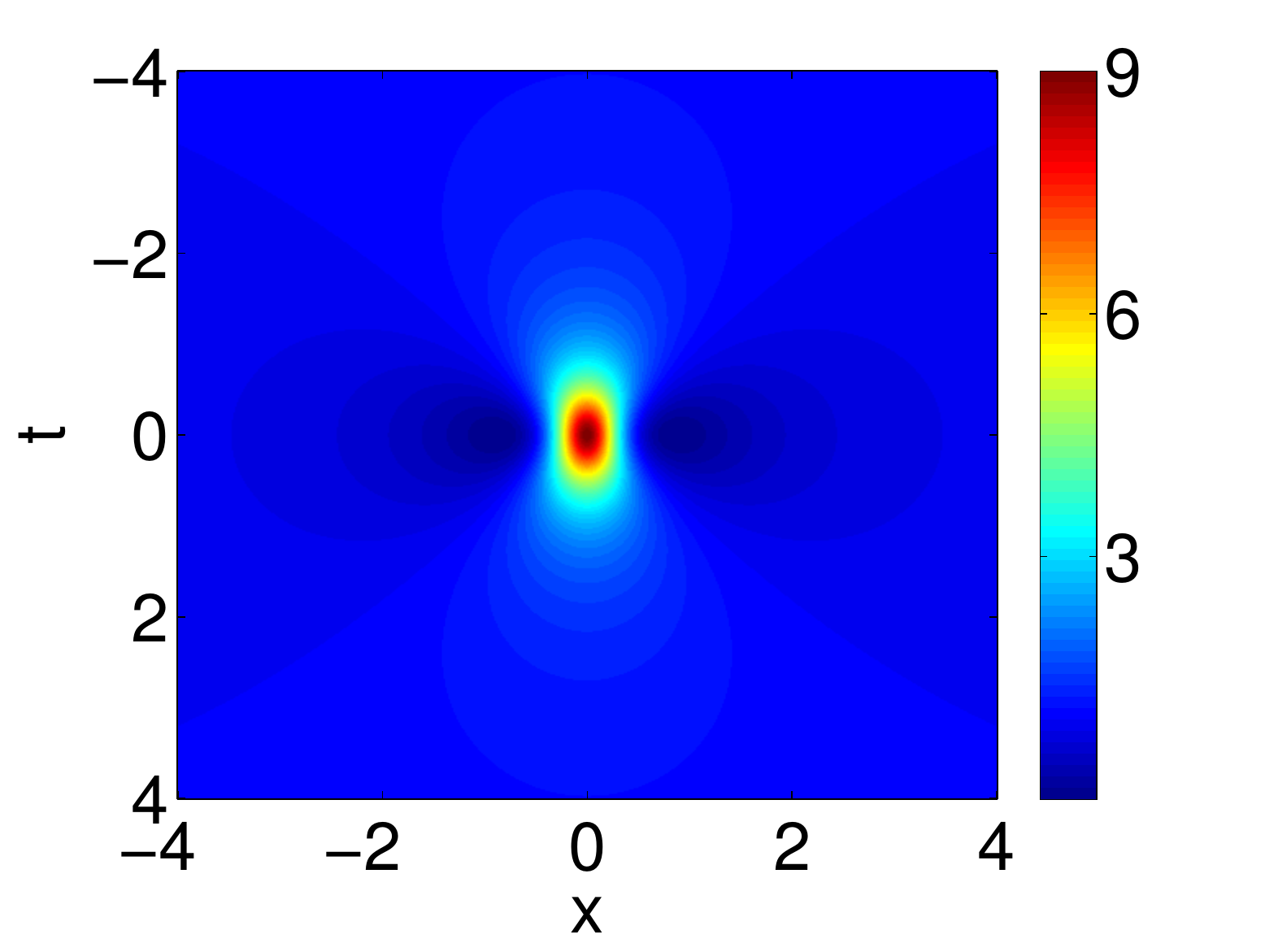}
      \includegraphics[height=0.28\textwidth]{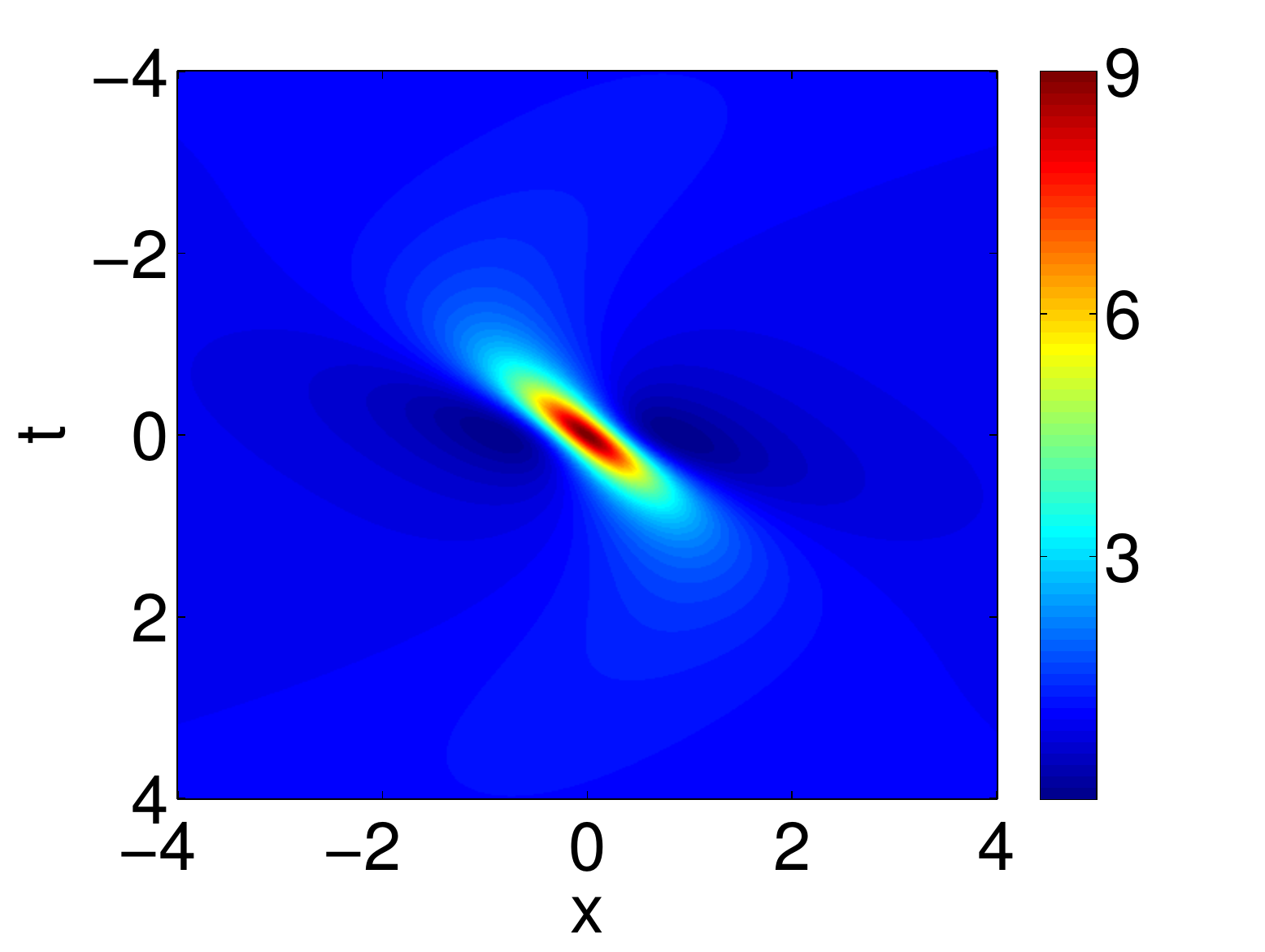}
\caption{Contour plots showing the density profiles of the rogue wave solution
of the NLS equation (left panel) and the rogue wave solution of Eq.~(\ref{model3}) (right panel).
Observe the relative ``twist'' of the latter about the origin $(0,0)$ in the $(x-t)$ plane.}
\label{Ffig3}
\end{figure}

Let us now proceed with the presentation of the rogue wave solution of Eq.~(\ref{model3}), which
has the form:
\begin{equation}
\psi= -\dfrac{
3-4x^2+8x\sin(t)-4\sin^{2}(t)-4t^2+8it}{
1+4x^2-8x\sin(t)+4\sin^{2}(t)+4t^2} e^{-(i/4)[-4x\cos(t)+\cos(t)\sin(t)-3t]}.
\label{r3}
\end{equation}
Notice that since $p=1$ in this case, the background of this rogue wave is constant.
In fact, it is found that the density profile of the above solution
bears many similarities with the one of the NLS equation [cf. Eq.~(\ref{rwofnlse})];
nevertheless, the orientation of the rogue wave is different in this case, as can be seen
by a ``twist'' about the origin $(0,0)$ in the $(x-t)$ plane --cf. Fig.~\ref{Ffig3}.
For this reason, hereafter, the rogue wave of Eq.~(\ref{r3}) will be called ``twisted rogue wave''.


\section{Numerical Results}

\subsection{Numerical methods.}
In this section, we present results of direct numerical integration of Eq.~(\ref{eqvcnlse}). 
While our analysis above indicates the existence of
{\it exact} solutions of Eq.~(\ref{eqvcnlse}), the time evolution dynamics
is used to explore the robustness of these explicit solutions under
the numerically induced perturbations (due not only to roundoff error,
but also due to the local truncation error arising by spatially and temporally
discretizing the dynamics).
Our numerical scheme is briefly described in what follows.

Initially, the second-order
partial derivatives with respect to $x$ are replaced by
second-order central differences on a grid consisting of $N$ equidistant points of
the form $x_{j}=-L+2jL/(N+1)$ with $j=1,\dots,N$ and $\Delta x=2L/(N+1)$, where $L$
determines the size of the spatial domain. In what follows, the grid size and its
resolution are chosen and fixed to be $L=300$ and $\Delta x=0.05$,
respectively (although, structurally, our results were not especially
sensitive to the particular choices).
As regards boundary conditions,
we note that we use no-flux conditions at the edges of the spatial
grid, namely $\partial_{x}\psi(x,t)|_{x=-L}=\partial_{x}\psi(x,t)|_{x=L}=0$. The latter
are replaced by forward and backward first-order finite difference formulas, respectively.

As far as the time integration is concerned, the Dormand and Prince (DOP853) method
with an automatic time-step procedure \cite{Hairer} is employed to solve the underlying
system of nonlinear ODEs. However, the standard 4th-order Runge-Kutta (RK4) method with
a fixed time-step is employed as well, and used to compare its results with the
ones found using the former method. Throughout the subsequent presentation, and for the RK4
method in particular, the time step-size ranges from $\Delta t=10^{-4}$ to $\Delta t=10^{-6}$
depending on the particular case studied. Finally, the initialization of the dynamics is performed
using the rogue wave solution of Eq.~(\ref{newslofvcnlse}), adjusted
to the models studied in the previous section. Notice that since in the
expression of Eq.~(\ref{newslofvcnlse}), the rogue wave arises around $t=0$,
the numerical integration is initialized shortly {\it before} $t=0$,
namely at a slightly negative time, in order to observe the
rogue wave formation, as it arises in our modified NLS models.

\subsection{Results of the simulations.}

\begin{figure}[t!]
\begin{center}
\vspace{-2.9cm}
\mbox{\hspace{-0.2cm}
\subfigure[][]{\hspace{-1.0cm}
\includegraphics[height=.21\textheight, angle =0]{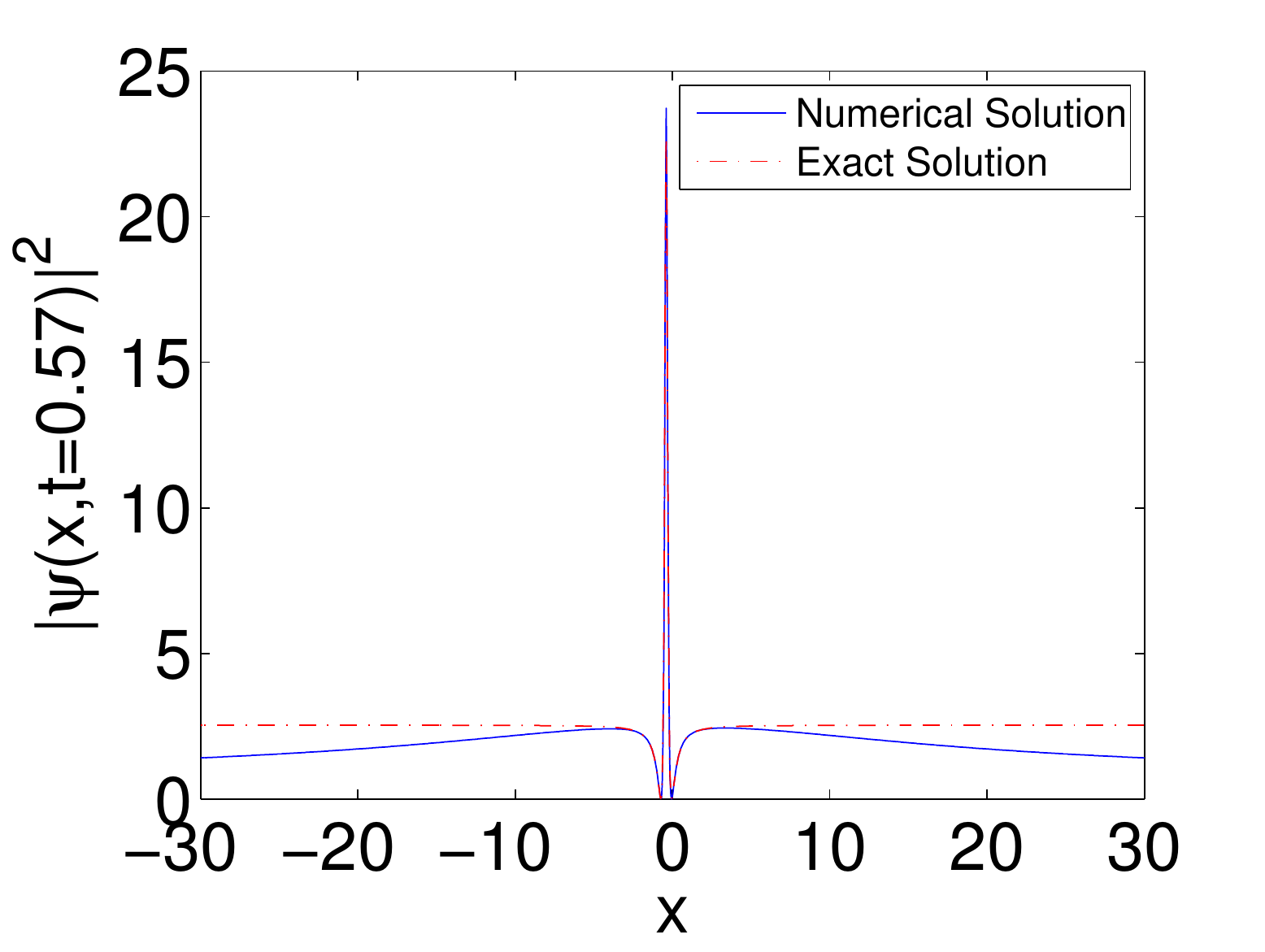}
\label{fig1a}
}
\subfigure[][]{\hspace{-0.5cm}
\includegraphics[height=.21\textheight, angle =0]{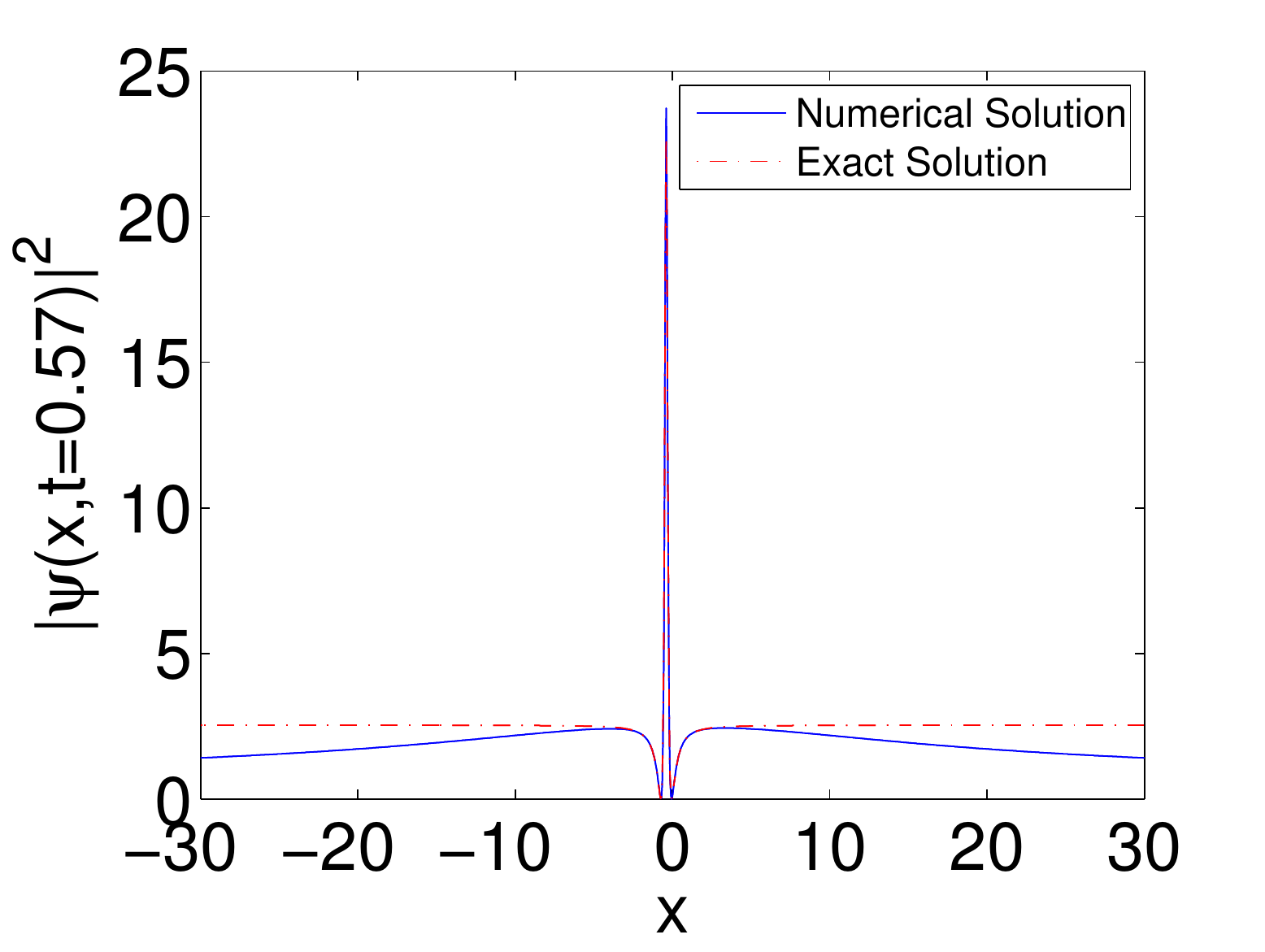}
\label{fig1b}
}
}
\mbox{\hspace{-0.2cm}
\subfigure[][]{\hspace{-1.0cm}
\includegraphics[height=.21\textheight, angle =0]{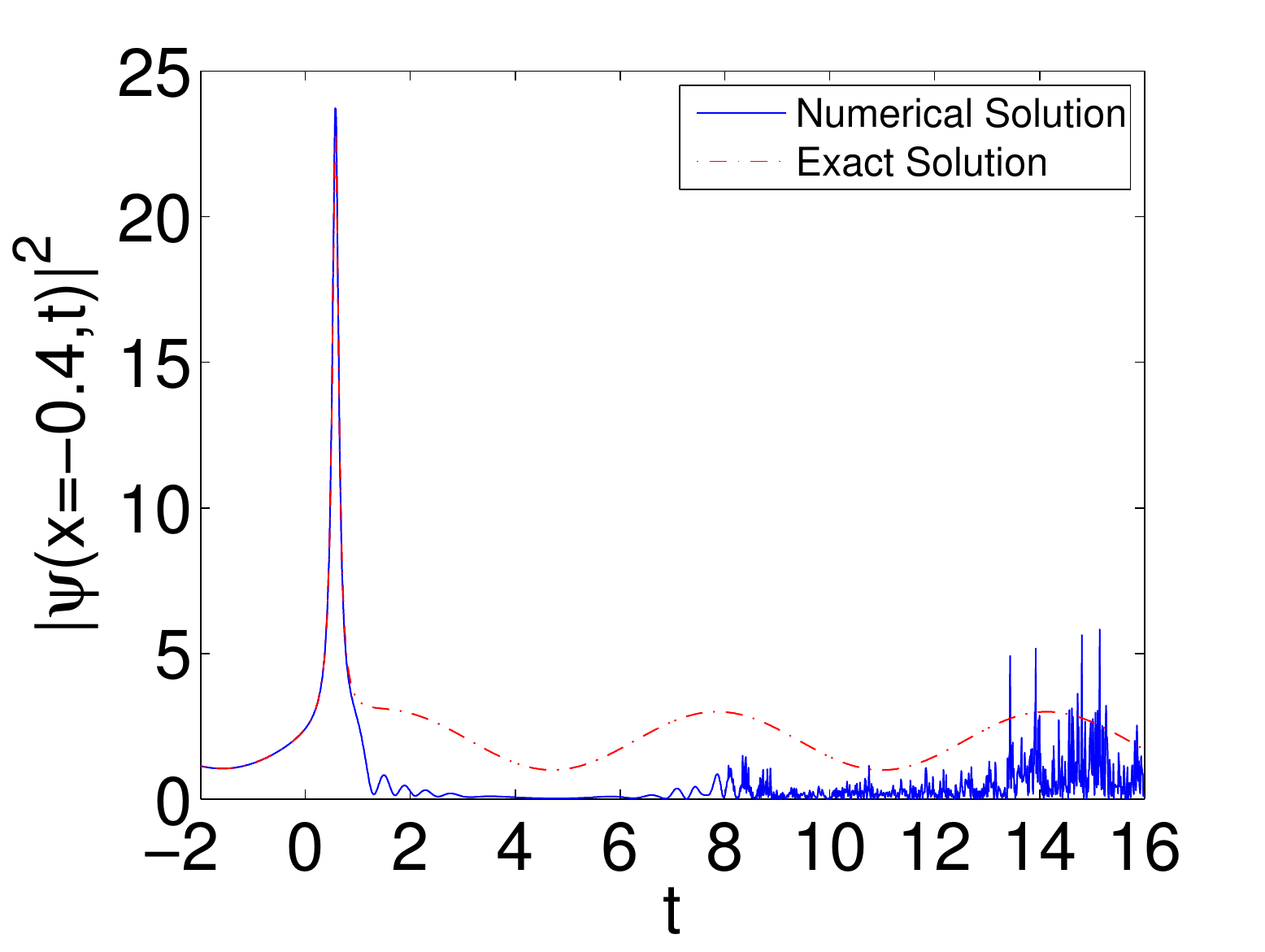}
\label{fig1c}
}
\subfigure[][]{\hspace{-0.5cm}
\includegraphics[height=.21\textheight, angle =0]{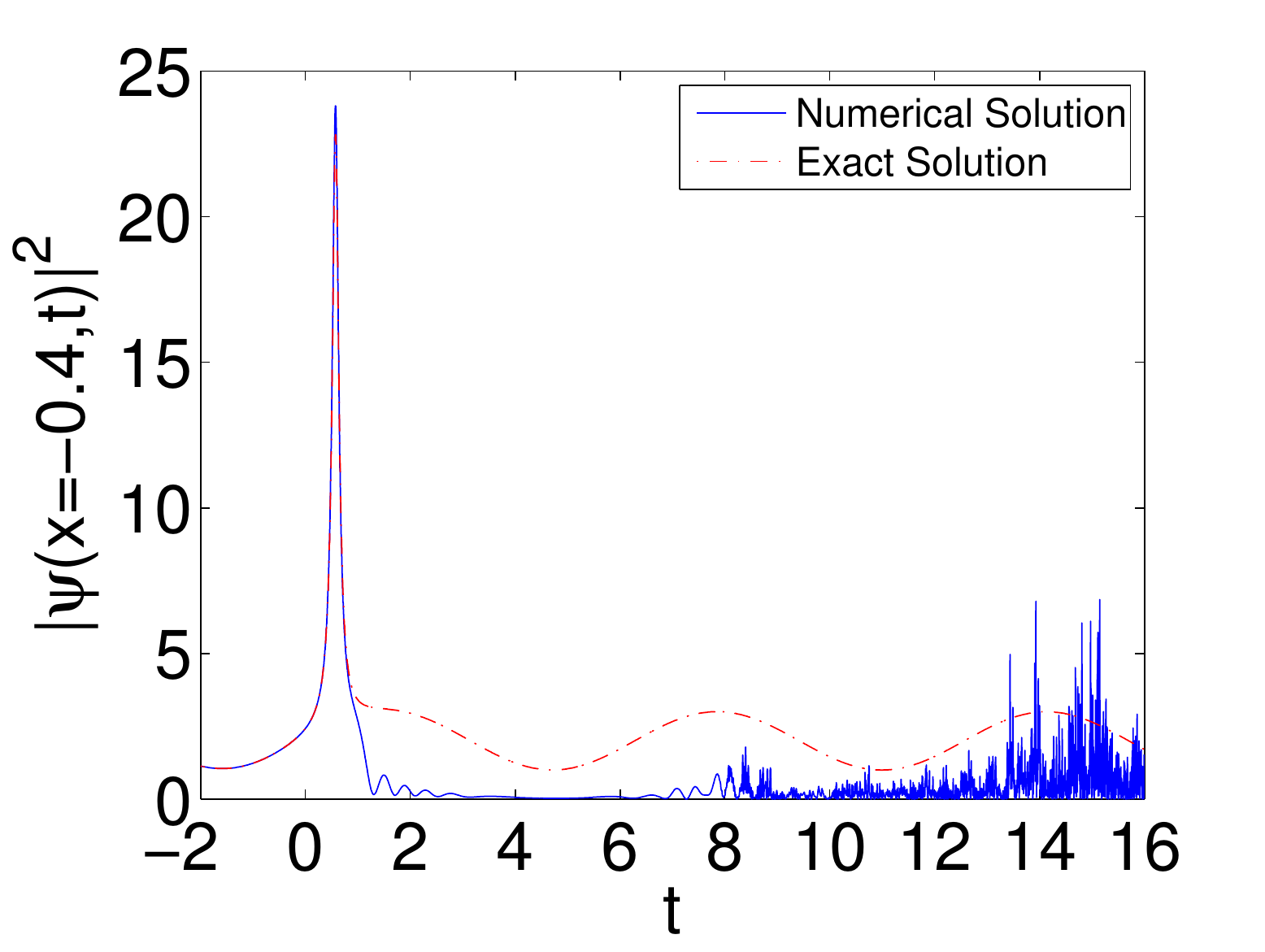}
\label{fig1d}
}
}
\mbox{\hspace{-0.2cm}
\subfigure[][]{\hspace{-1.0cm}
\includegraphics[height=.21\textheight, angle =0]{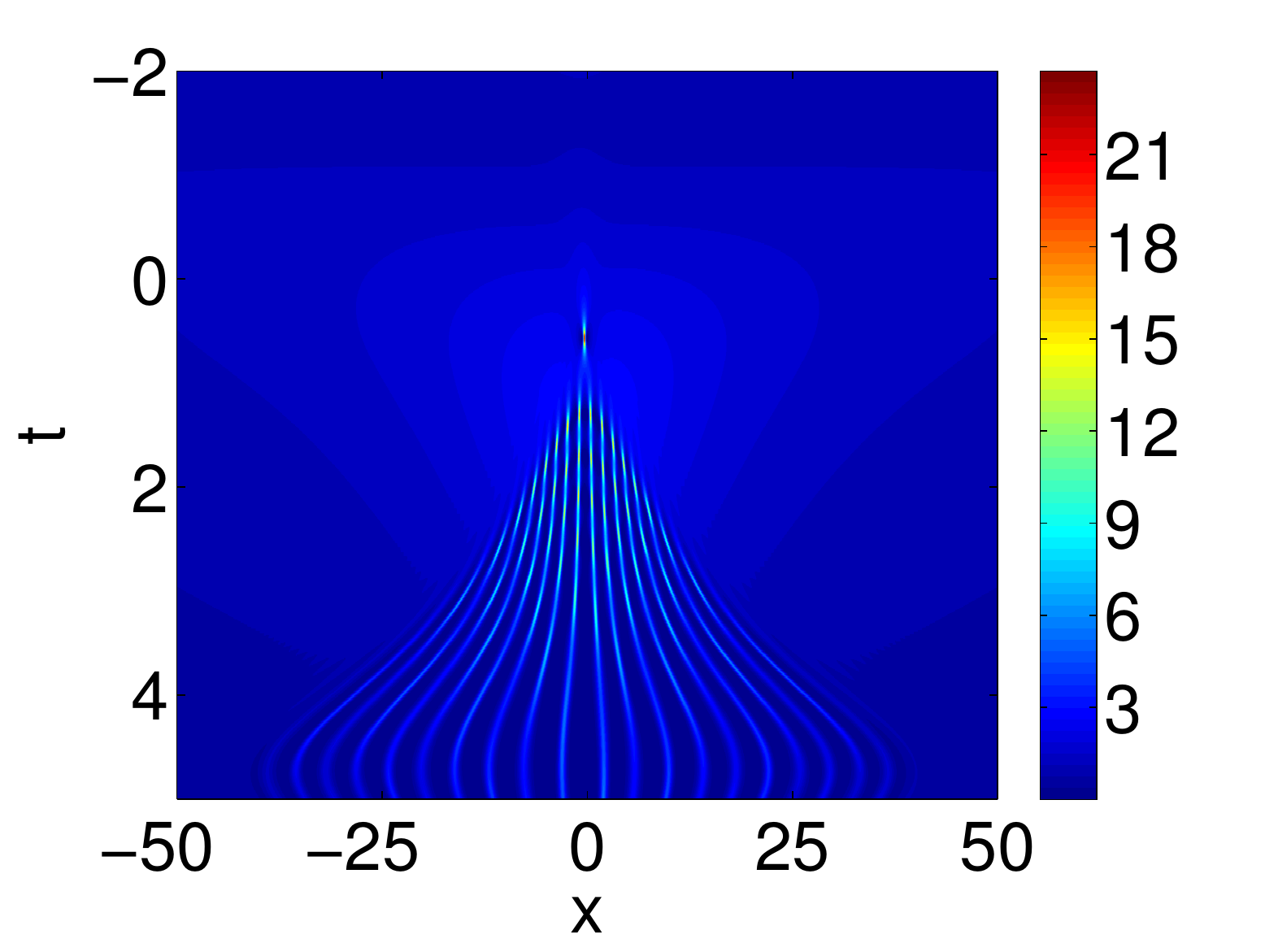}
\label{fig1e}
}
\subfigure[][]{\hspace{-0.5cm}
\includegraphics[height=.21\textheight, angle =0]{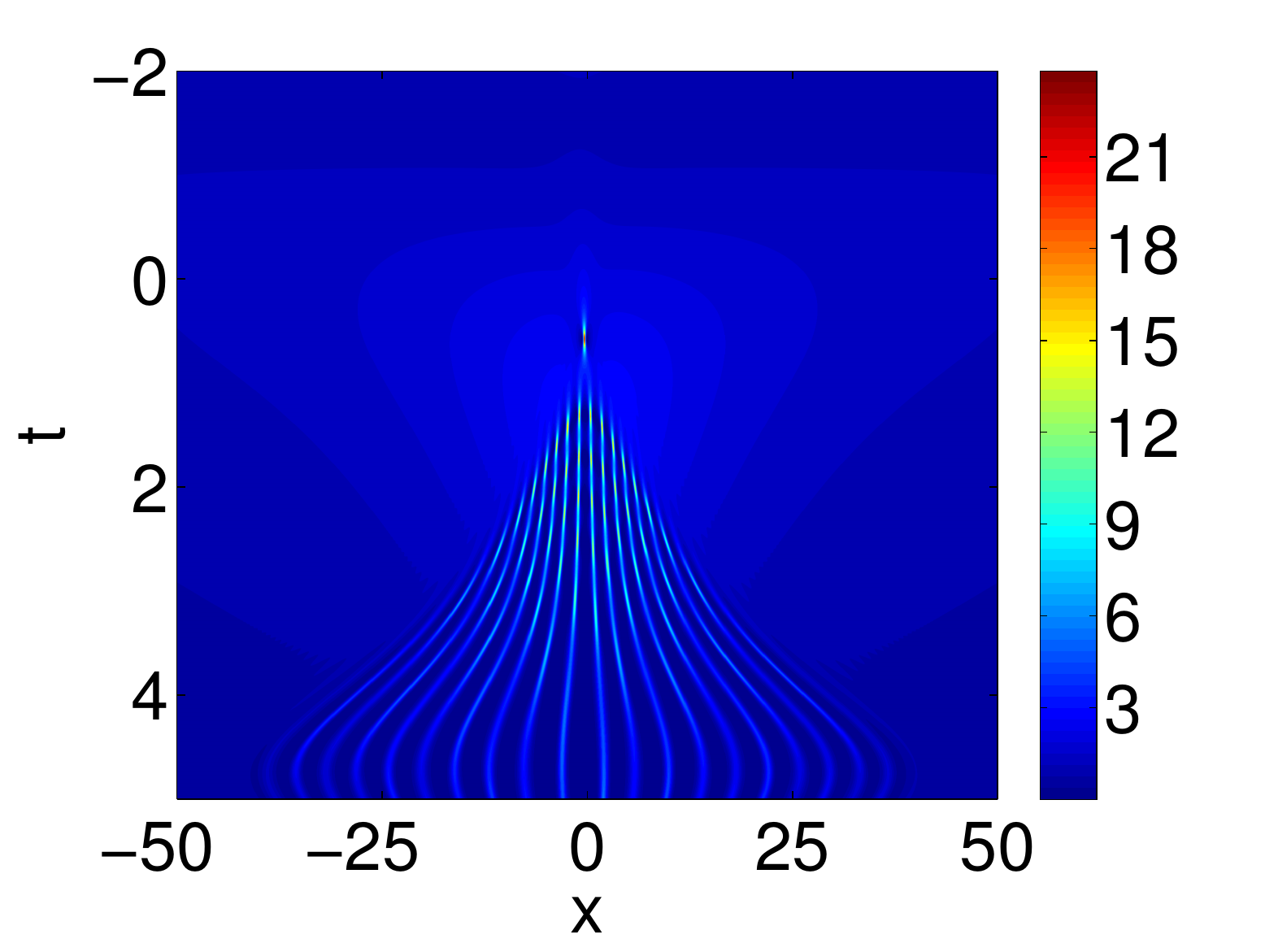}
\label{fig1f}
}
}
\mbox{\hspace{-0.2cm}
\subfigure[][]{\hspace{-1.0cm}
\includegraphics[height=.21\textheight, angle =0]{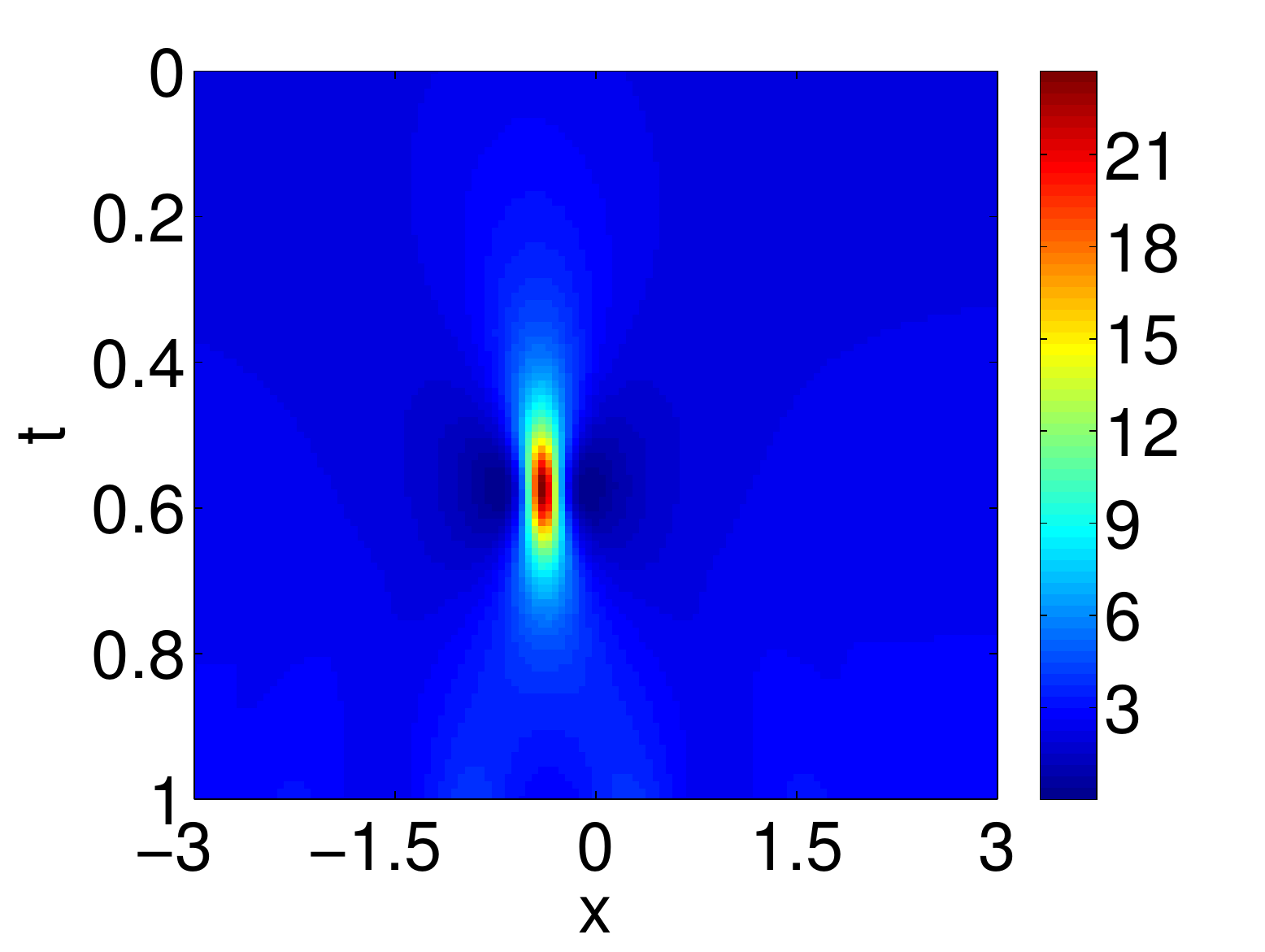}
\label{fig1g}
}
\subfigure[][]{\hspace{-0.5cm}
\includegraphics[height=.21\textheight, angle =0]{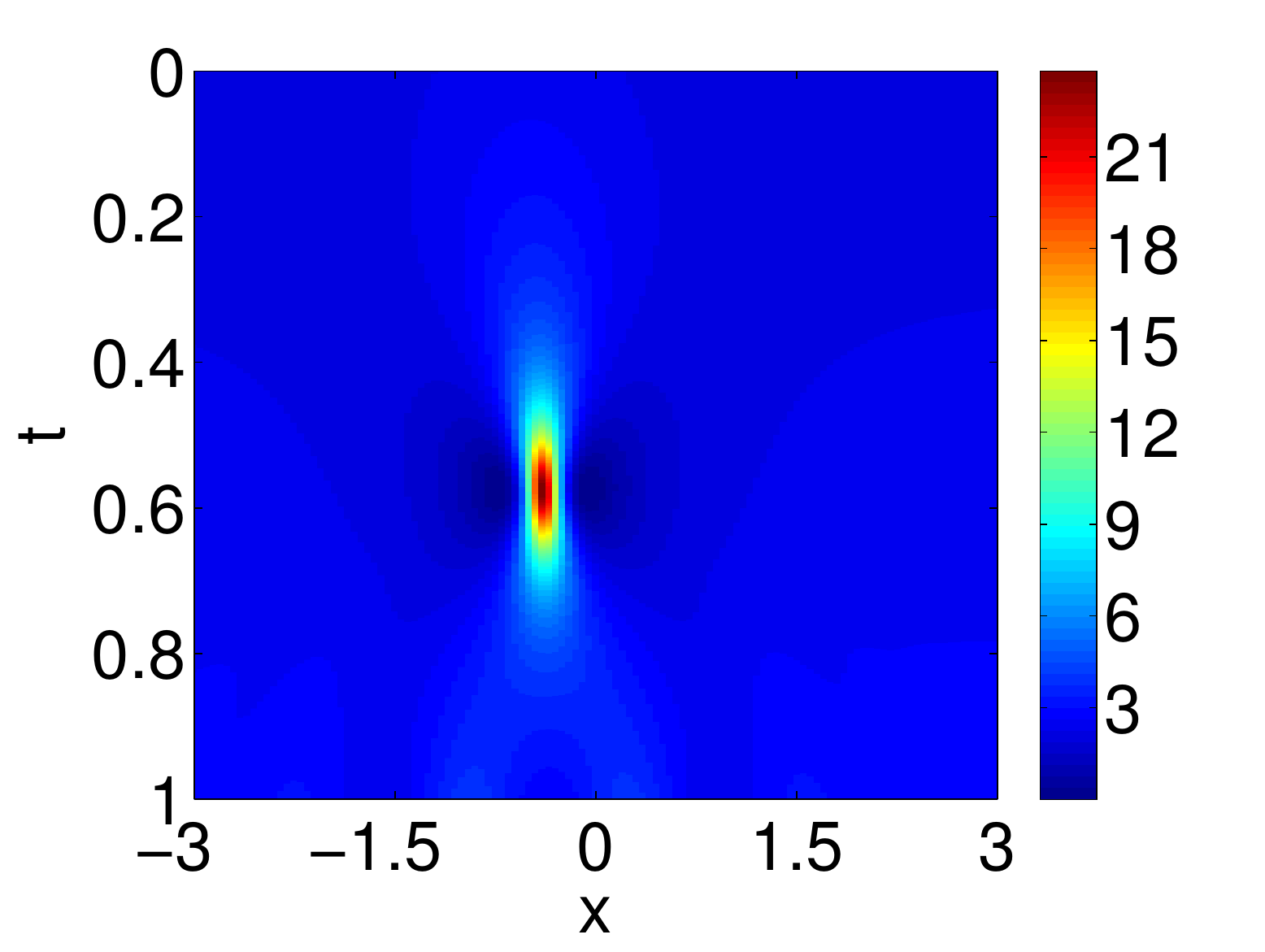}
\label{fig1h}
}
}
\end{center}
\caption{
Rogue wave on a periodic background. Left and right panels correspond to numerical results
using the DOP853 method and the RK4 one, respectively. The top panels (a-b) show the spatial
distribution of the intensity $|\psi|^{2}$ at $t=0.57$, whereas the panels (c-d) show its
temporal evolution 
at $x=-0.4$. The panels (e-h) show contour plots of the density
profile of the rogue wave (a wider and a narrower view of the field around
the rogue wave is provided).}
\label{fig1}
\end{figure}

\begin{figure}[t!]
\begin{center}
\vspace{-2.9cm}
\mbox{\hspace{-0.2cm}
\subfigure[][]{\hspace{-1.0cm}
\includegraphics[height=.21\textheight, angle =0]{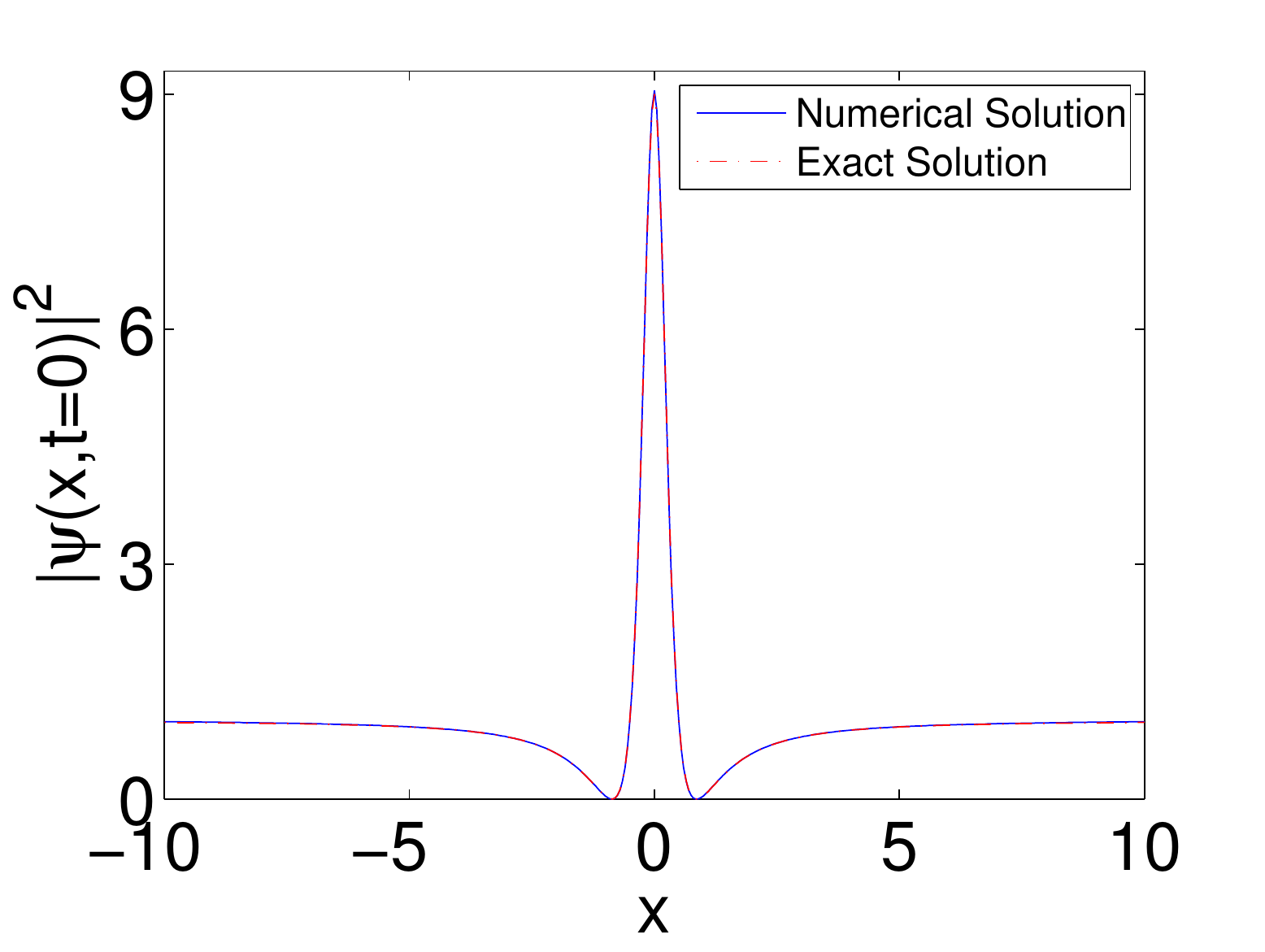}
\label{fig2a}
}
\subfigure[][]{\hspace{-0.5cm}
\includegraphics[height=.21\textheight, angle =0]{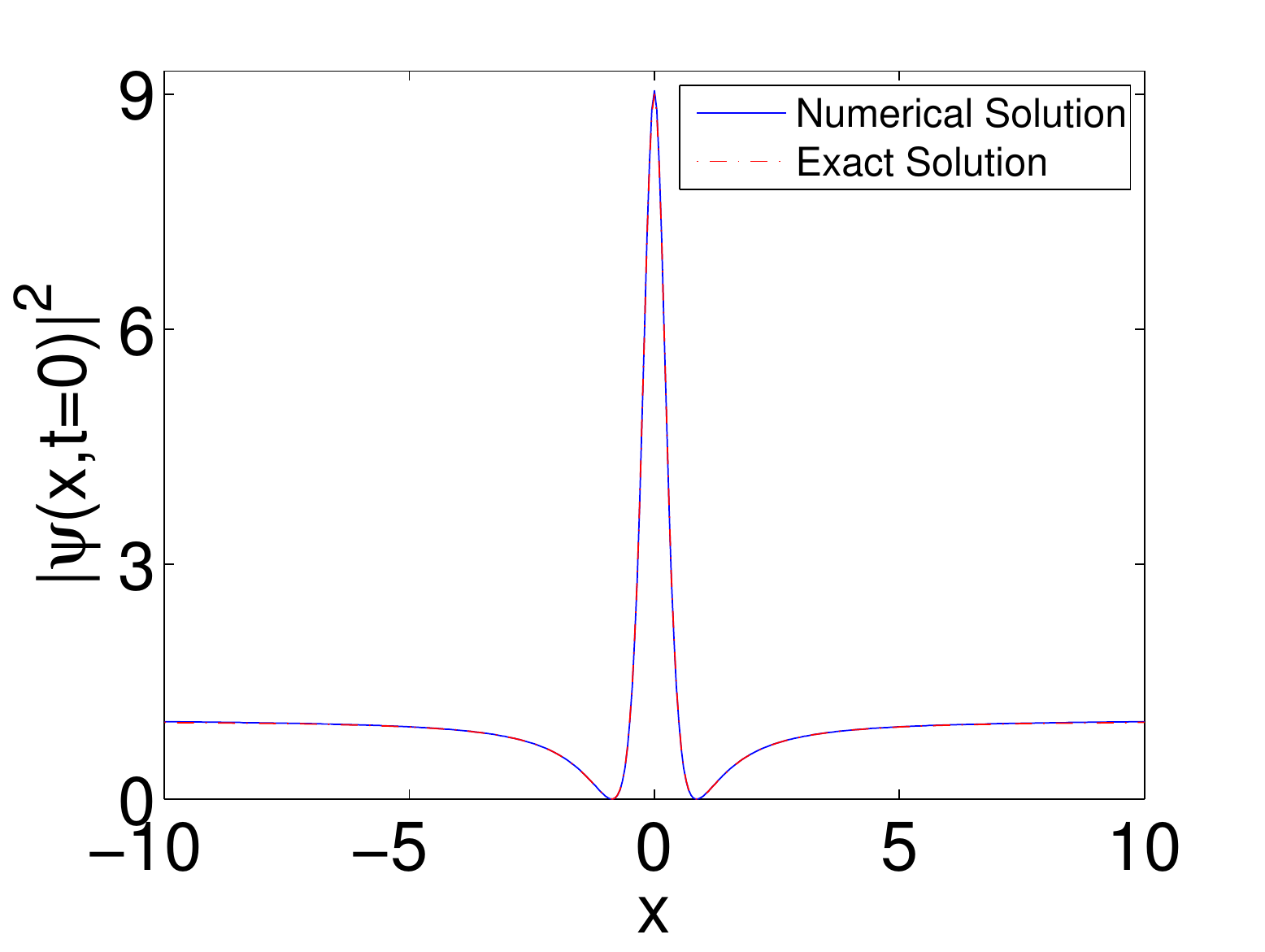}
\label{fig2b}
}
}
\mbox{\hspace{-0.2cm}
\subfigure[][]{\hspace{-1.0cm}
\includegraphics[height=.21\textheight, angle =0]{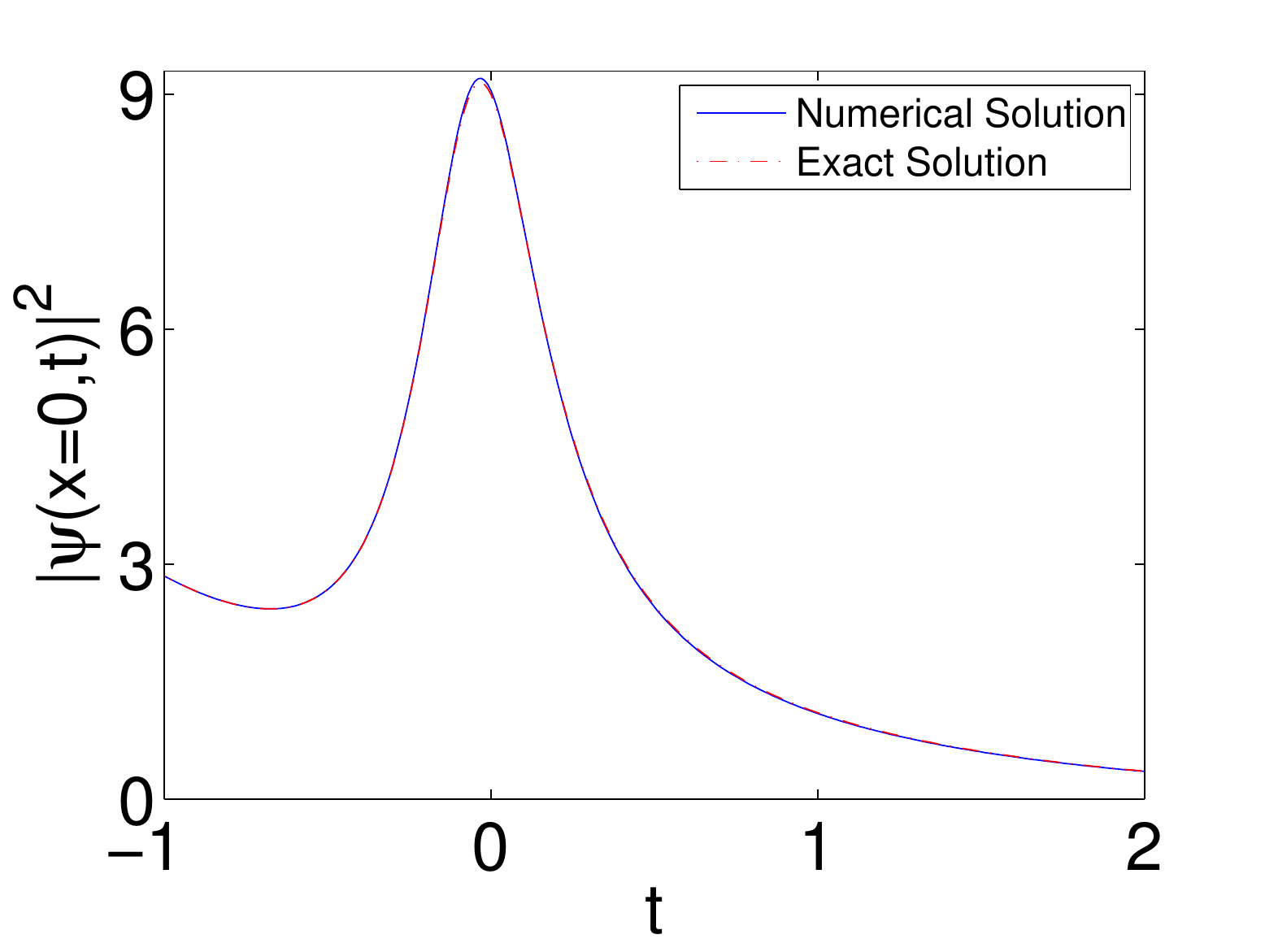}
\label{fig2c}
}
\subfigure[][]{\hspace{-0.5cm}
\includegraphics[height=.21\textheight, angle =0]{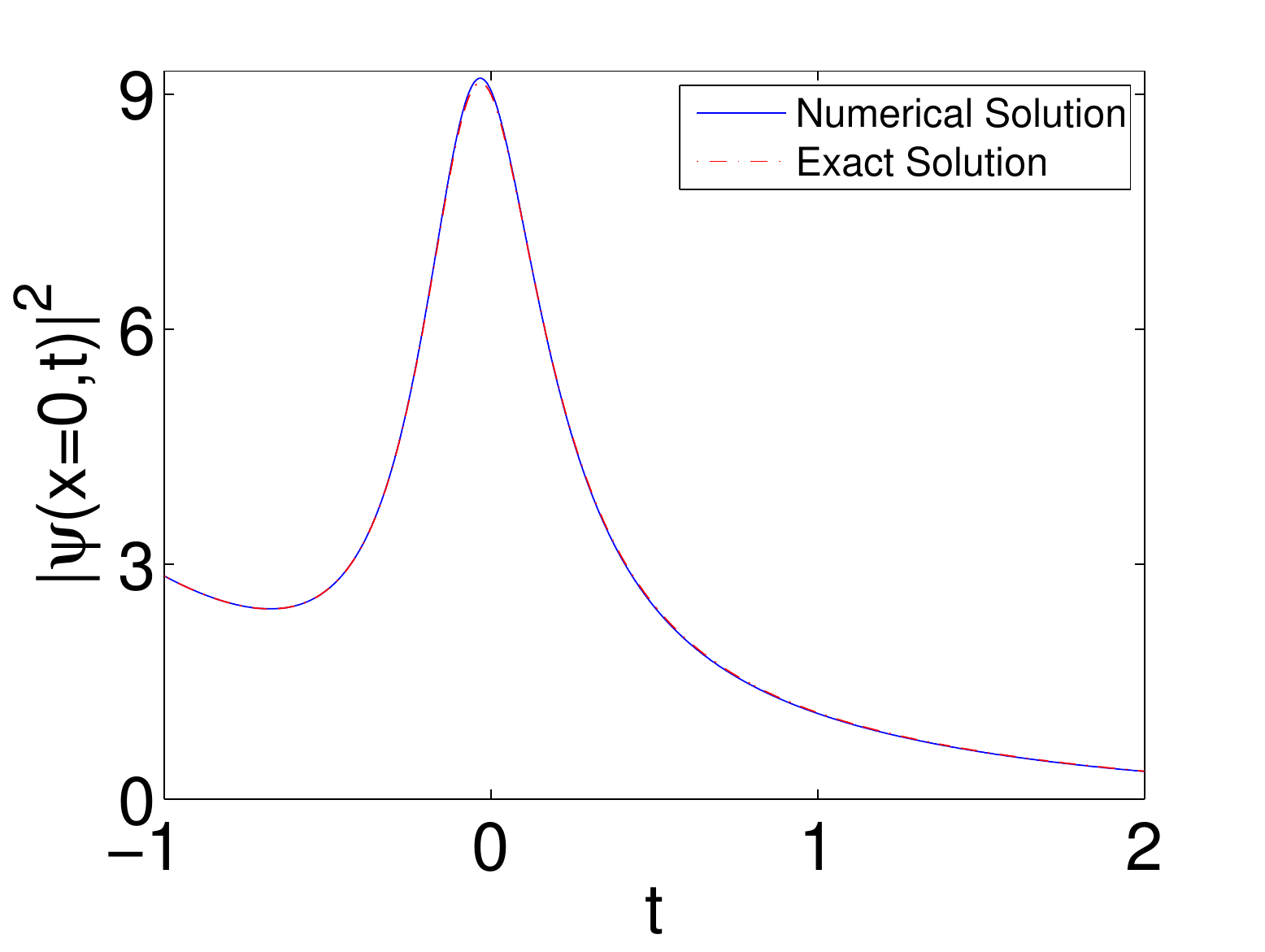}
\label{fig2d}
}
}
\mbox{\hspace{-0.2cm}
\subfigure[][]{\hspace{-1.0cm}
\includegraphics[height=.21\textheight, angle =0]{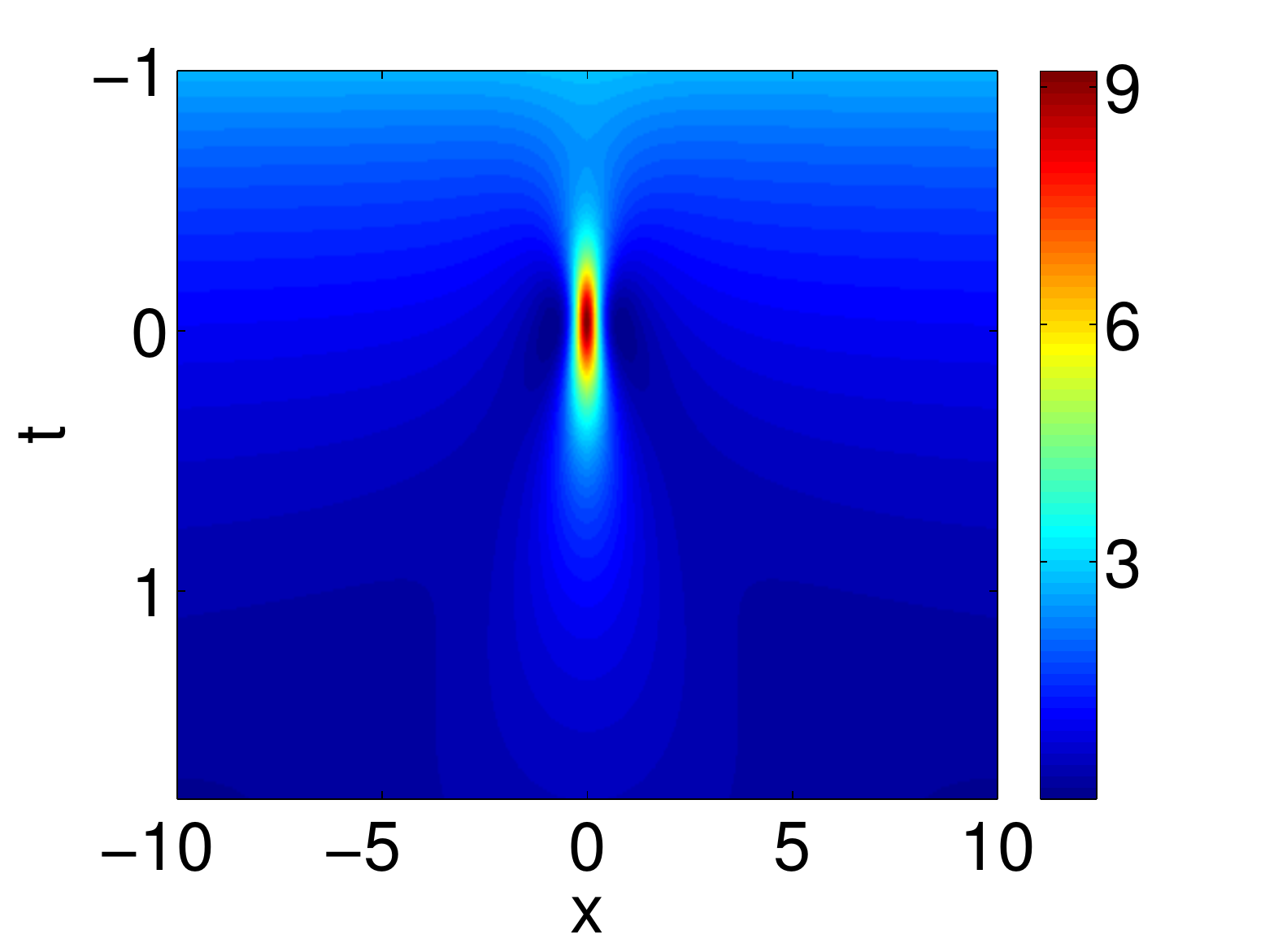}
\label{fig2e}
}
\subfigure[][]{\hspace{-0.5cm}
\includegraphics[height=.21\textheight, angle =0]{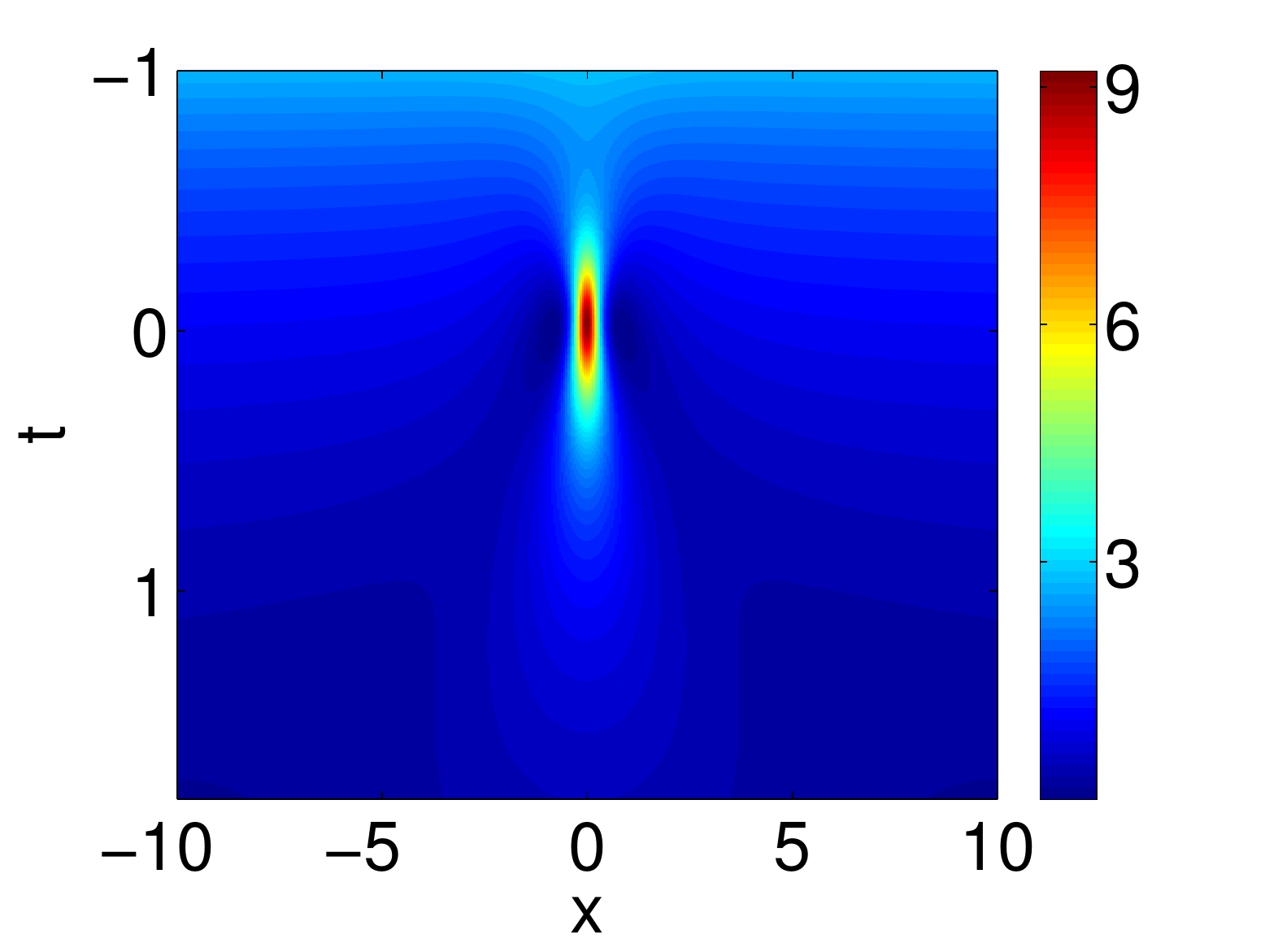}
\label{fig2f}
}
}
\mbox{\hspace{-0.2cm}
\subfigure[][]{\hspace{-1.0cm}
\includegraphics[height=.21\textheight, angle =0]{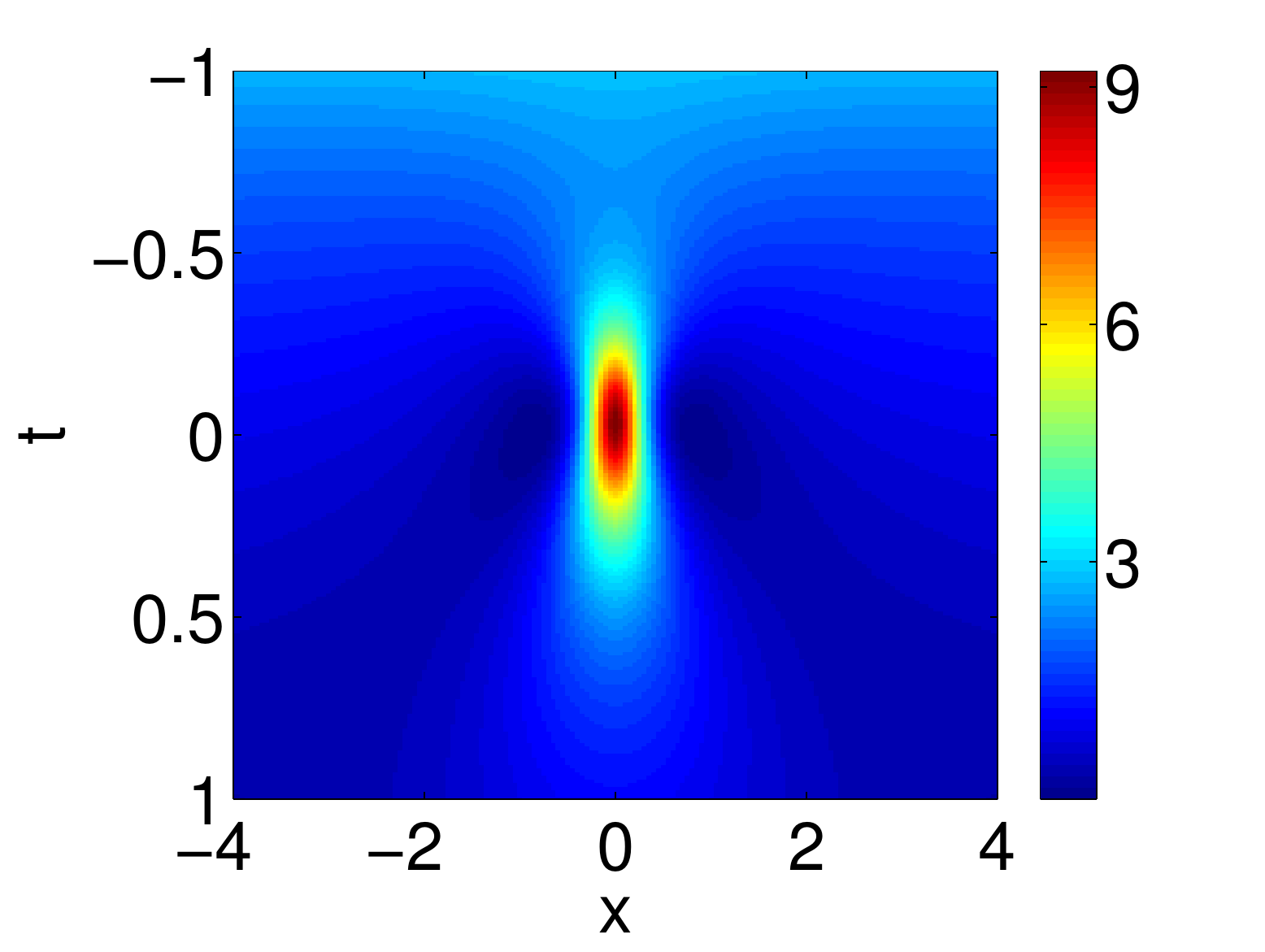}
\label{fig2g}
}
\subfigure[][]{\hspace{-0.5cm}
\includegraphics[height=.21\textheight, angle =0]{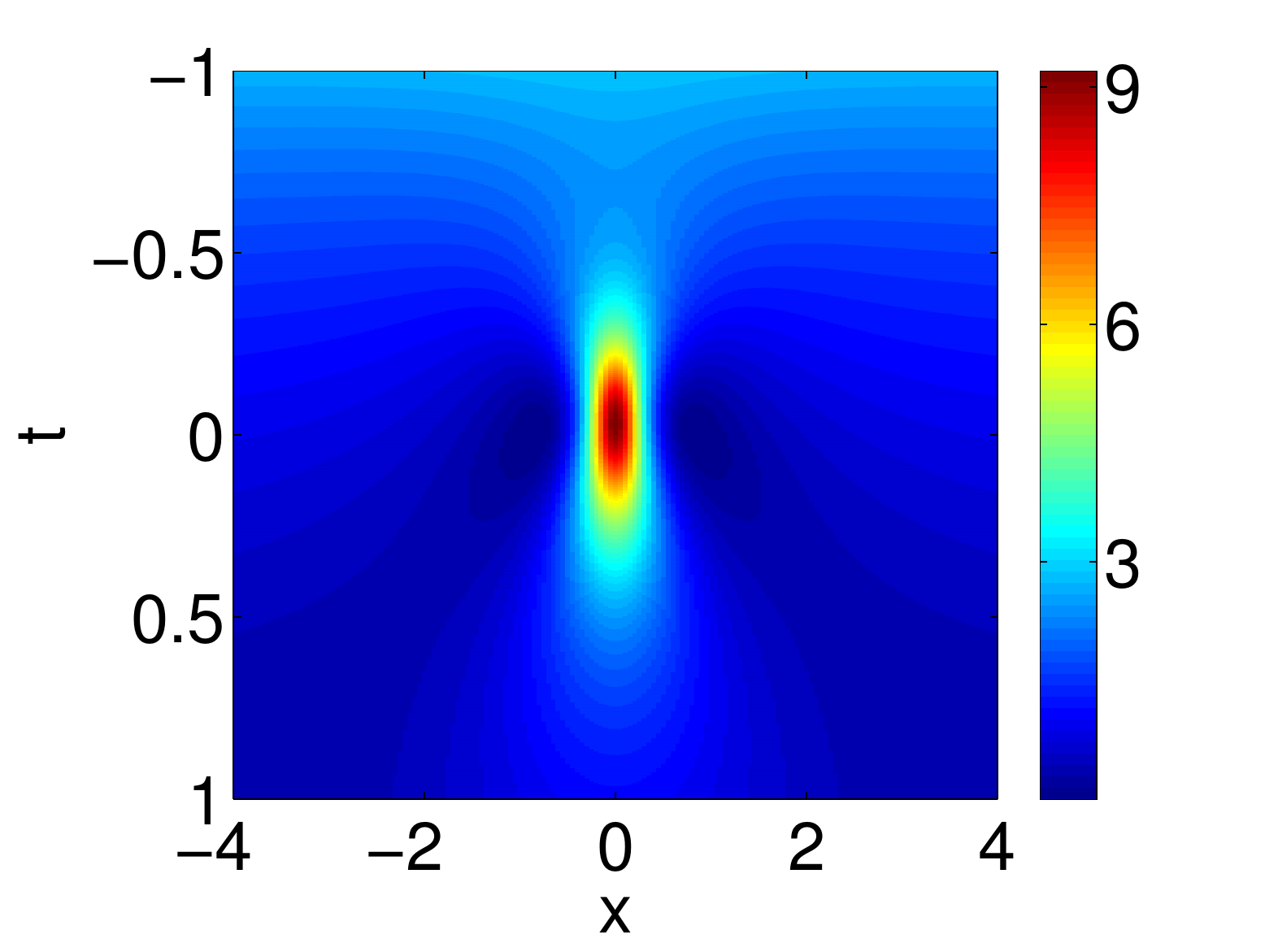}
\label{fig2h}
}
}
\end{center}
\caption{
Rogue wave on a monotonically decreasing background. Left and right panels correspond to
numerical results using the DOP853 method and the RK4 one, respectively. The top panels (a-b)
show the spatial distribution of the intensity $|\psi|^{2}$ at $t=0$, whereas the panels
(c-d) show its temporal evolution 
at $x=0$. The panels (e-h) show contour plots of
the density profile of the rogue wave (again, presenting a wider and a
narrower space-time view of the density field).}
\label{fig2}
\end{figure}

\begin{figure}[t!]
\begin{center}
\vspace{-2.9cm}
\mbox{\hspace{-0.2cm}
\subfigure[][]{\hspace{-1.0cm}
\includegraphics[height=.21\textheight, angle =0]{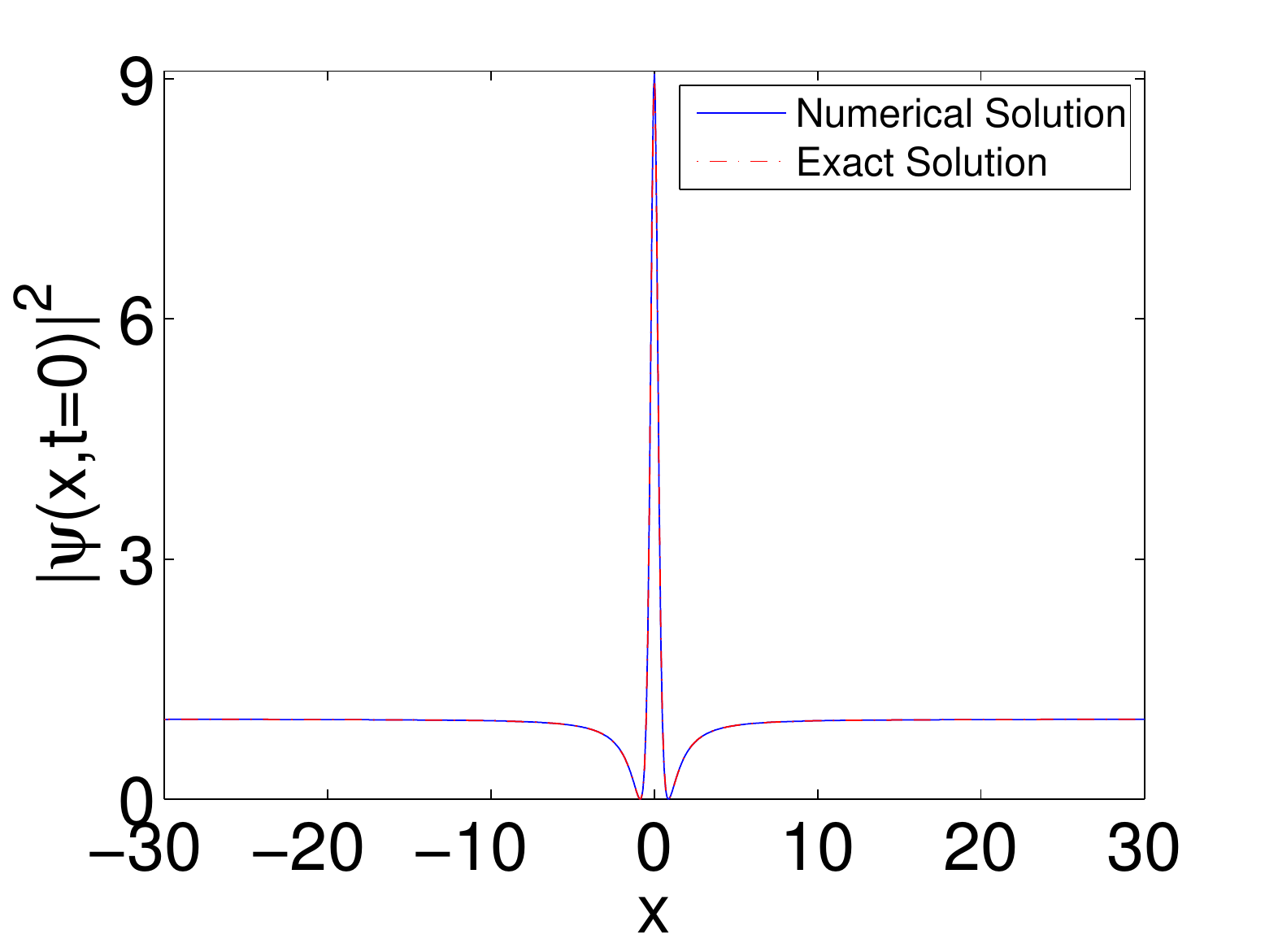}
\label{fig3a}
}
\subfigure[][]{\hspace{-0.5cm}
\includegraphics[height=.21\textheight, angle =0]{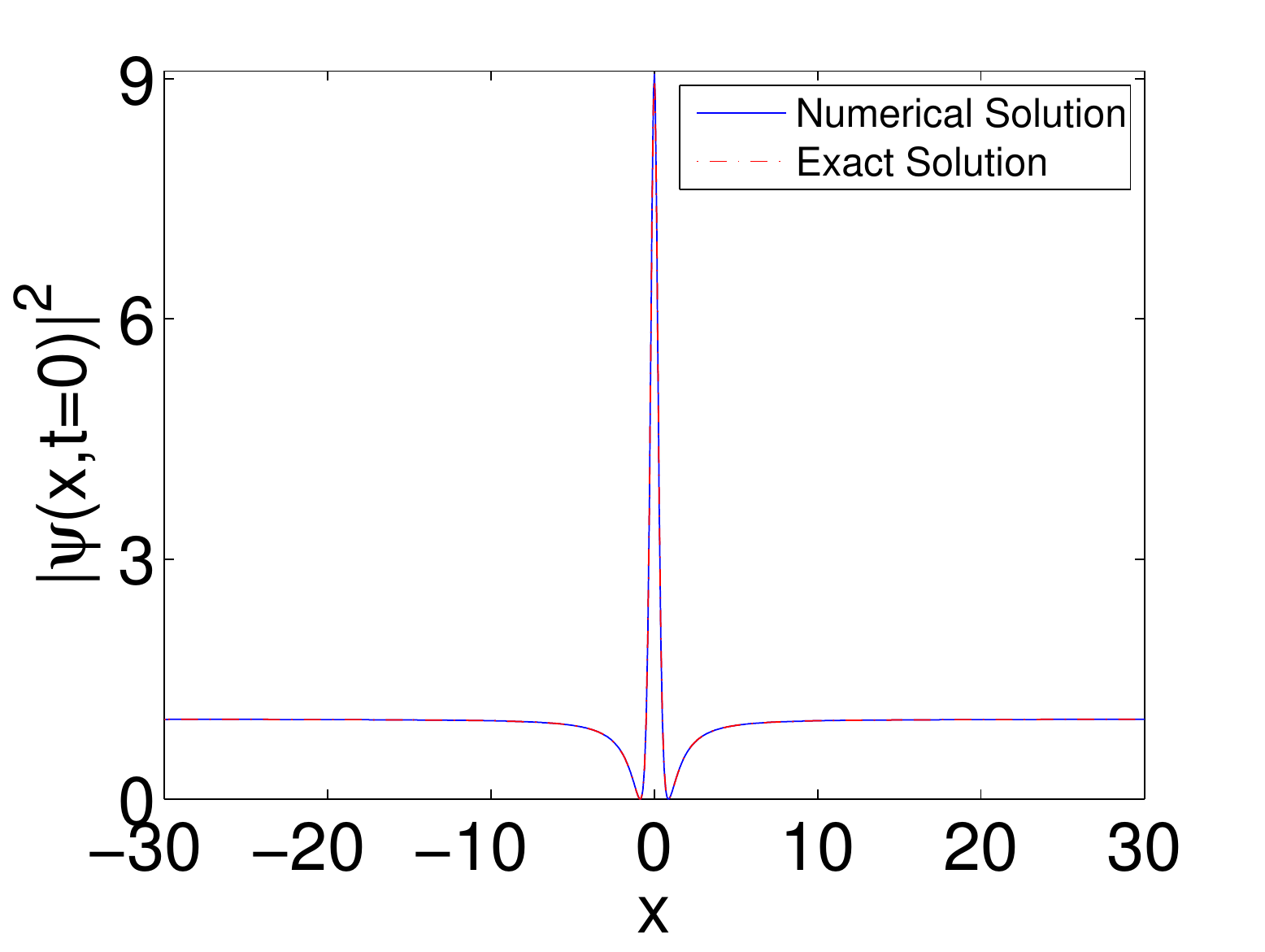}
\label{fig3b}
}
}
\mbox{\hspace{-0.2cm}
\subfigure[][]{\hspace{-1.0cm}
\includegraphics[height=.21\textheight, angle =0]{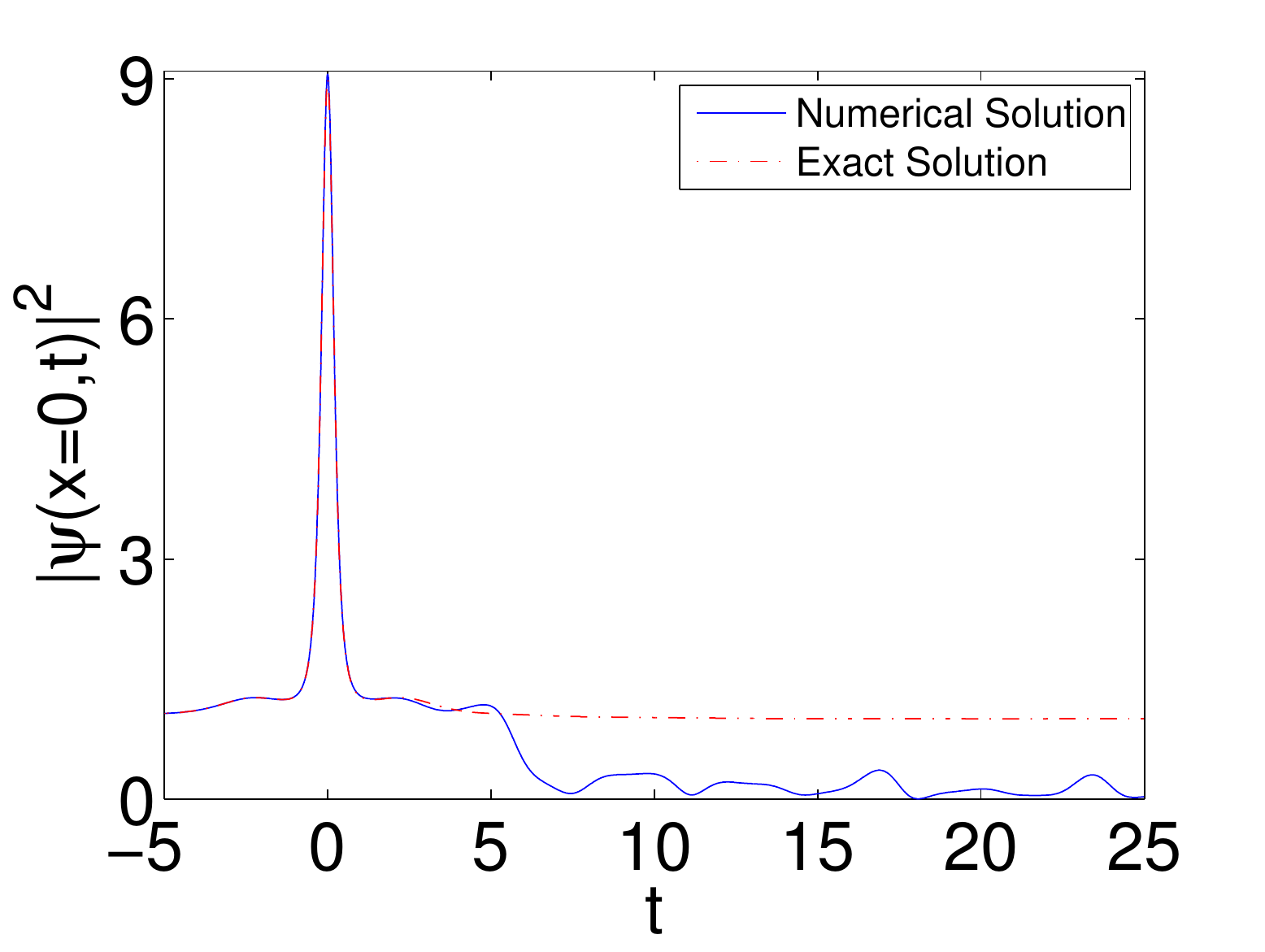}
\label{fig3c}
}
\subfigure[][]{\hspace{-0.5cm}
\includegraphics[height=.21\textheight, angle =0]{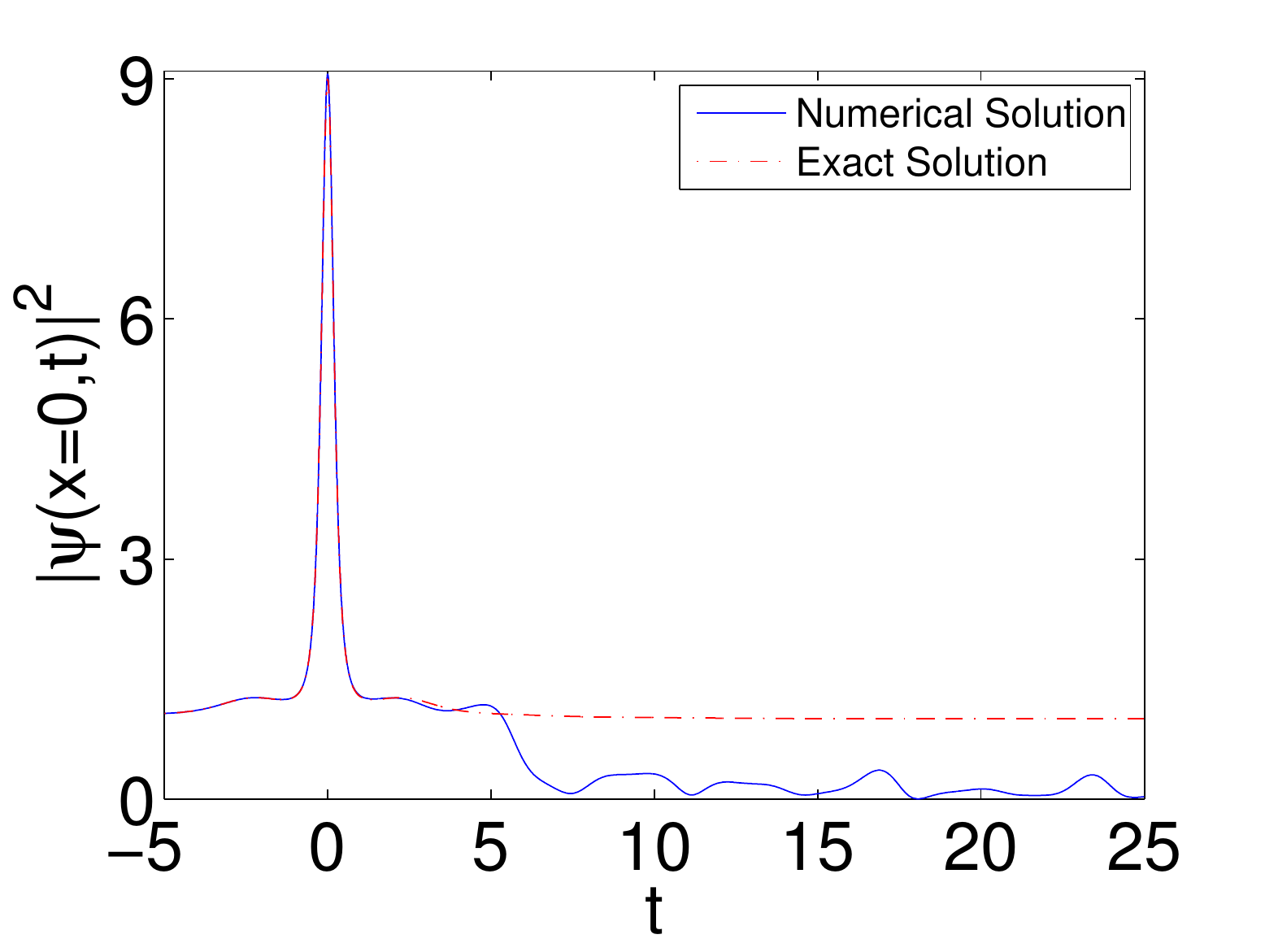}
\label{fig3d}
}
}
\mbox{\hspace{-0.2cm}
\subfigure[][]{\hspace{-1.0cm}
\includegraphics[height=.21\textheight, angle =0]{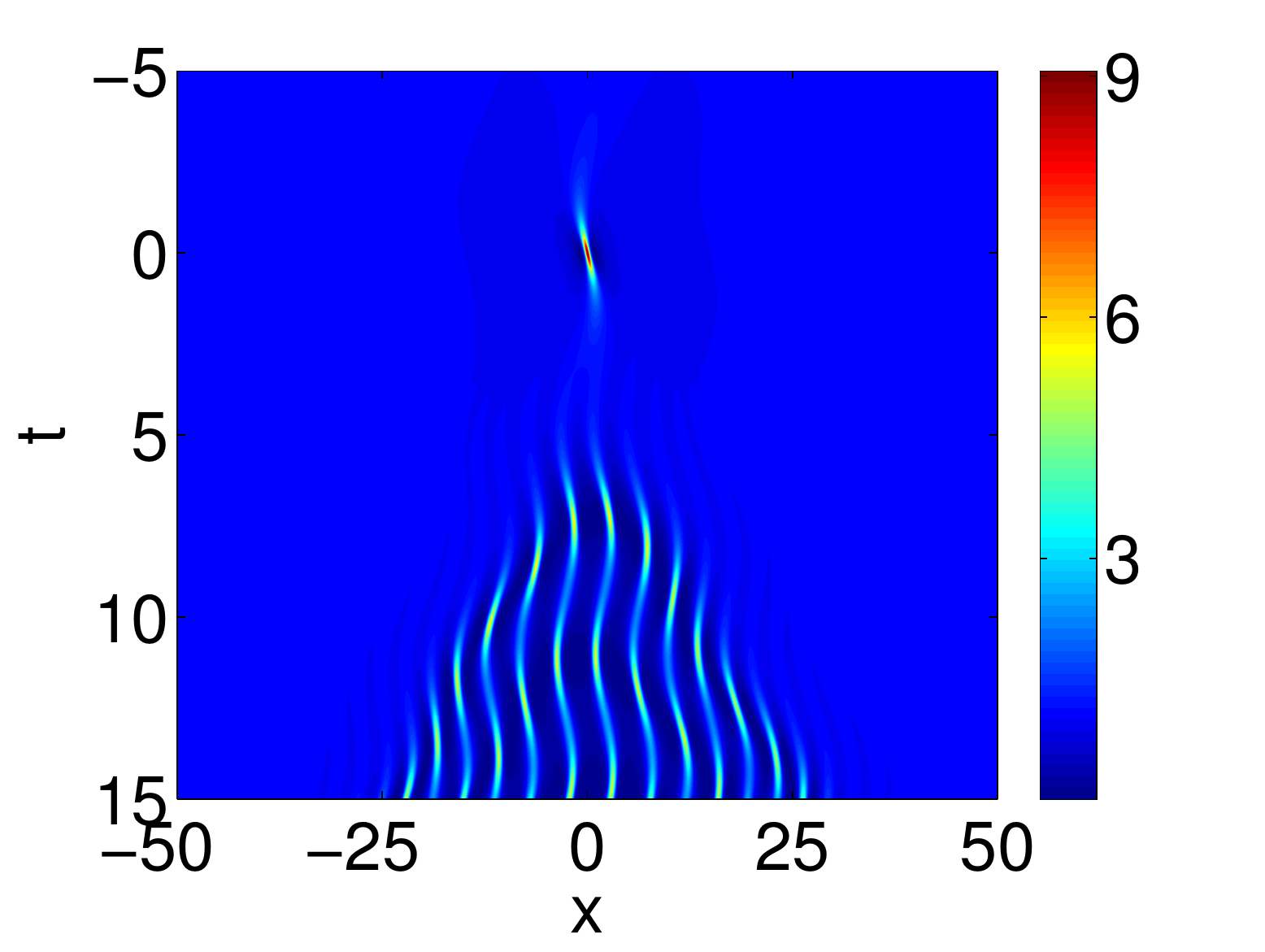}
\label{fig3e}
}
\subfigure[][]{\hspace{-0.5cm}
\includegraphics[height=.21\textheight, angle =0]{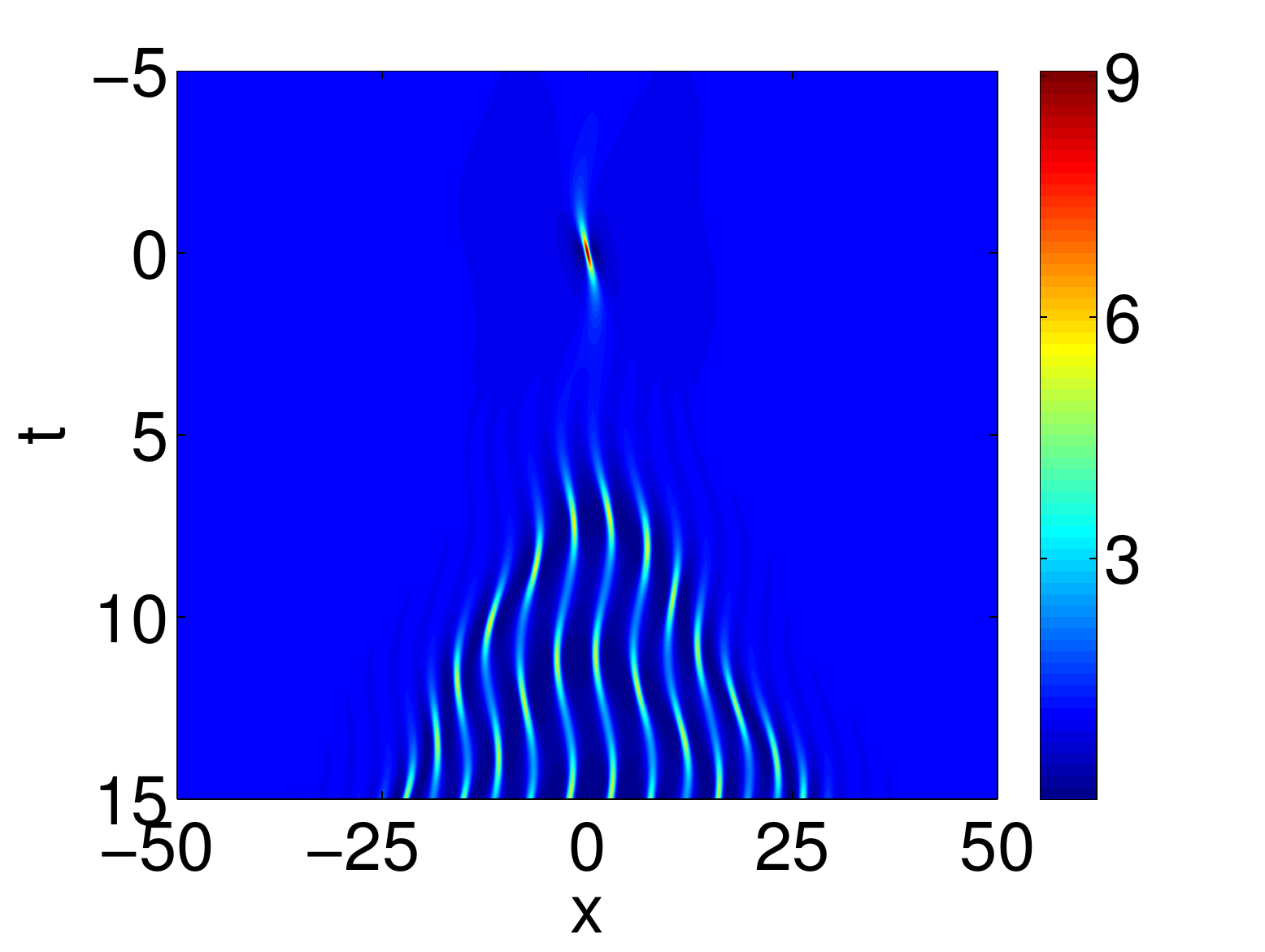}
\label{fig3f}
}
}
\mbox{\hspace{-0.2cm}
\subfigure[][]{\hspace{-1.0cm}
\includegraphics[height=.21\textheight, angle =0]{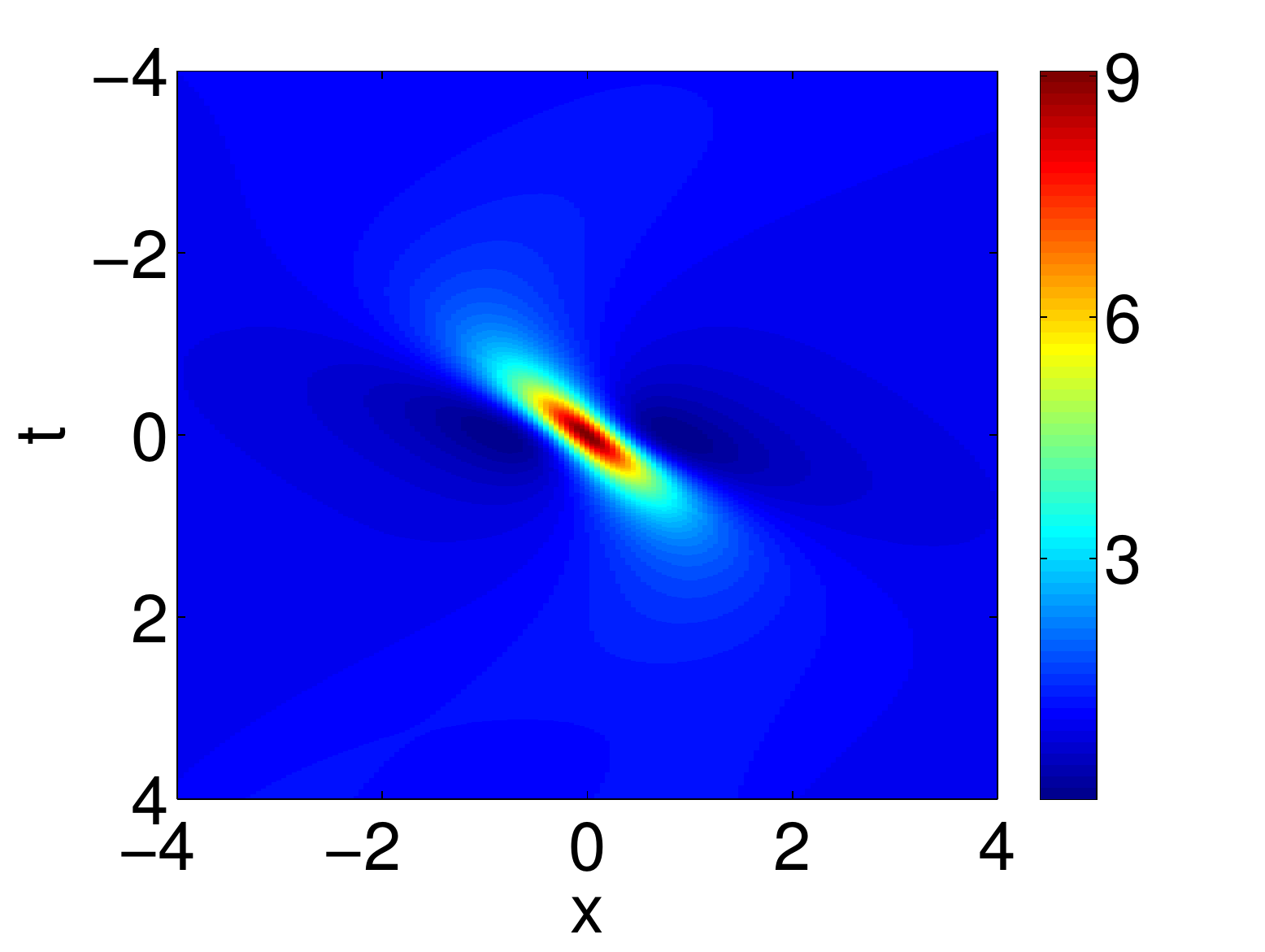}
\label{fig3g}
}
\subfigure[][]{\hspace{-0.5cm}
\includegraphics[height=.21\textheight, angle =0]{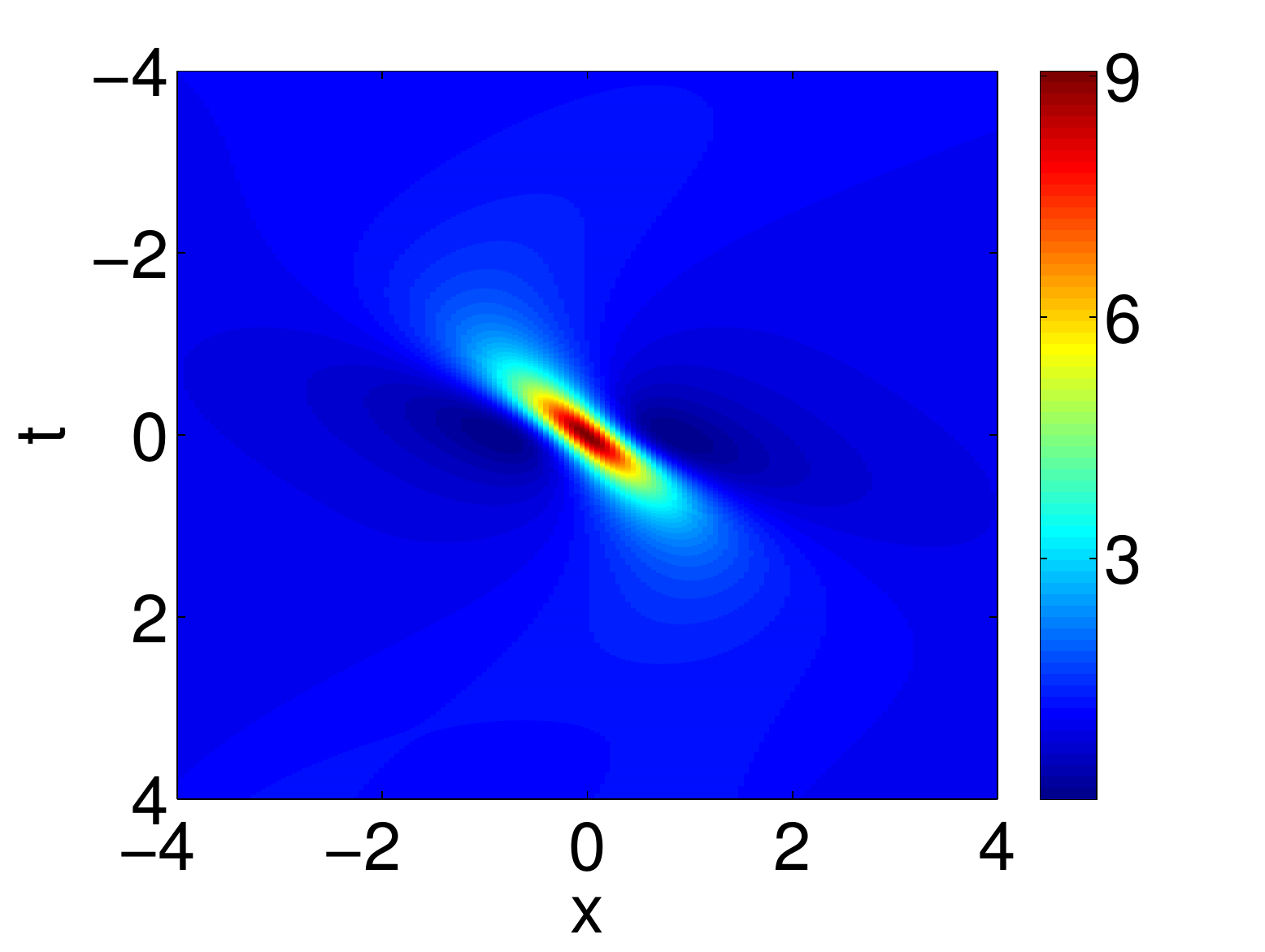}
\label{fig3h}
}
}
\end{center}
\caption{
Twisted rogue wave. Left and right panels correspond to numerical results using the DOP853 method
and the RK4 one, respectively. The top panels (a-b) show the spatial distribution of the intensity
$|\psi|^{2}$ at $t=0$, whereas the panels (c-d) show its temporal evolution at $x=0$. Finally,
the panels (e-h) show contour plots of the density profile of the rogue wave.}
\label{fig3}
\end{figure}

Having presented details of our numerical techniques,
we next present the results of our direct simulations for
the rogue wave solutions of Eq.~(\ref{eqvcnlse}). In particular, Figs.~\ref{fig1}, \ref{fig2}
and \ref{fig3} correspond, respectively,
to the rogue wave on a periodic background,
rogue wave on a monotonically decreasing background, and the twisted rogue wave case.
The respective time intervals $[t_{i},t_{f}]$ for the numerical integration (forward in time),
as well as the time step-sizes (for the RK4 method only) in each case are 
$[-2,16]$ with $\Delta t=10^{-5}$, $[-1,2]$ with $\Delta t=10^{-6}$ and $[-5,25]$ with $\Delta t=10^{-4}$. Note that the
left and right panels show results using the DOP853 and RK4 method
respectively, whereas
in panels (a-d), we also compare the numerical solutions (in particular their densities
$|\psi|^{2}$) with the exact ones (see Section~2).
The latter are denoted by a dashed
red line. Finally, the panels (e-h) complement our results by
illustrating space-time contour plots
for the density profile of each of the rogue wave solutions (as identified
numerically).

These results suggest a number of interesting observations. On the one
hand, in all cases, the rogue wave formation is clearly observed
and the dynamics of our high order integrators up to the time
of its emergence seems to quite accurately capture the expected
behavior on the basis of the identified exact analytical solution.
Remarkably though, in Figs.~\ref{fig1} and~\ref{fig3} and despite
the fact that our system was initialized with the exact solitary
wave rogue state, beyond the occurrence time of the large amplitude
structure, the dynamics starts being quite different than the one
expected theoretically (analytically). This, in turn, suggests
that the small, numerical roundoff error and/or the local
truncation error in numerically approximating our partial
differential equation is being strongly amplified as the solution
reaches these especially large amplitudes. As a result, the
evolution past the formation of the rogue wave presents a
radically different spatio-temporal shape than the homogeneous
one expected on the basis of the Peregrine soliton dynamics.
More specifically, what we see as a result is the clear formation
in both Figs.~\ref{fig1} and~\ref{fig3} of an emerging {\it modulational
instability} that rapidly converts the homogeneous state (that
was supposed to be observed for positive times,
past the formation of the Peregrine soliton) into a large
array of solitary waves, arising in a highly ordered pattern.
This evolutional outcome is perhaps not particularly surprising
in light of the well known modulational instability of the relevant
uniform state. However, it is perhaps partially intriguing to
further explore in light of the experimental observability of
the Peregrine soliton (which would be normally taken to suggest
a significant structural robustness of such a solution, which,
however, we note as being apparently absent here).

It is interesting to note that while the rogue wave on a periodic
background and the twisted one both develop such a progressively
expanding pattern
of solitary waves, the latter feature is not observed in our
second example of a monotonically decreasing background. However,
this can be  understood too, as the modulational
instability growth rate is well known to monotonically depend on the
amplitude of the background wave. Hence, in this case of decreasing
background wave amplitude, so does the corresponding growth rate, not
allowing the instability to be observed in the
time window of interest herein.

\clearpage

\clearpage

\section{Conclusions}
In conclusion, we explored some case examples of rogue wave solutions
in NLS systems with variable coefficients. Starting from a general analytical framework, and motivated by the
possibility of atomic physics applications in the realm of Bose-Einstein condensates, we focused on
NLS models with constant dispersion. Through  the choice of arbitrary functions and parameters involved in
a transformation of the unknown field, and the use of a ``seed solution'' of an auxiliary NLS model 
(in proper coordinates), in the form of a rogue wave, we were led to three different NLS models that we
argued as being relevant to  the physics of atomic BECs. These models, in particular, may
be used to describe BEC settings where the external potential or/and the atomic scattering length (i.e., the
nonlinearity coefficient in the NLS model) depend on time. We discussed variants of these 
possibilities that have already been implemented in actual BEC experiments
and which suggest the realizability of our proposed models.

For the above mentioned three different models, exact analytical rogue wave solutions were
determined. 
Our analytical results were also tested against direct numerical simulations.
In particular, these solutions were employed as initial conditions for the direct numerical integration
of the original NLS equations. We used two different integration schemes,
based on the
Dormand and Prince (DOP853) method, and the more standard 4th-order
Runge-Kutta (RK4) method.
The numerical results we obtained by these two methods
clearly demonstrated the formation of the rogue waves in close
agreement with
the analytical results (in both our numerical methods).

This agreement, together with the physical relevance of the considered
models, suggest that the
predicted rogue waves may have a good chance to be observed within the
current experimental capability of BEC setups.
That being said, we also observed some very interesting side-products
of the Peregrine solitary wave initial data. In particular, the apparent
amplification of local truncation errors, especially given the large
amplitude of the associated solutions led to significant deviations
from the uniform state anticipated (by means of the exact solution)
past the formation of the rogue wave in two out of three of our examples.
More specifically, the uniform state became subject to a modulational
instability initiated exactly at the location of the rogue wave
(where the error was apparently amplified maximally). This led
to the emergence of an ordered pattern of solitary waves which
subsequently expanded spatially as time evolved. This phenomenon
was intriguing in its own right and such a nonlinear evolution
of small perturbations to the Peregrine profile merits additional
investigation both from a numerical but perhaps also from an analytical
perspective (at the level of the regular NLS model). This is a natural
direction for further studies.

There are many additional directions that would also be relevant
and timely.  Here,
we focused on single Peregrine solitons, while other rogue wave
solutions (including periodic ones)
and a detailed comparison of their dynamics, would be particularly interesting. Additionally,
we explored systems with constant dispersion; nevertheless, systems with
varying dispersion
that are particularly relevant in the context of optics, would be another
theme for future  investigations. Finally, a potential generalization
of the present settings to higher dimensional configurations would be
particularly relevant to consider in numerous physical settings,
including fluid and superfluid, as well as nonlinear optical systems.
Such studies are currently in progress and will be reported in future
publications.


\vspace{0.5cm}
{\bf Note added.}
After the submission of this paper, we were informed of the following relevant works
on the modulation of breathers and localized solutions in single and multi-component
nonlinear Schr\"odinger equations: \cite{cardoso1}-\cite{cardoso3}.

\vspace{0.5cm}
{\bf Acknowledgments.}
This work is supported by the NSF of China under Grant No. 11271210 and the K. C. Wong Magna Fund
in Ningbo University. The work of D.J.F. was partially supported
by the Special Account for Research Grants of the University of Athens. E.G.C. is indebted
to both the Institute of Physics, Carl von Ossietzky University (Oldenburg) and the University of
Massachusetts (Amherst) for the kind hospitality provided where part of this work was carried
out there. He also gratefully acknowledges financial support from the German Research Foundation
DFG, the DFG Research Training Group 1620 ``{\it Models of Gravity}" and FP7 People IRSES-606096:
``{\it Topological Solitons, from Field Theory to Cosmos}".
P.G.K. also acknowledges support from the National Science Foundation
under grants CMMI-1000337, DMS-1312856, from the Binational Science
Foundation under grant 2010239, from FP7-People under grant IRSES-606096
and from the US-AFOSR under grant
FA9550-12-10332.



\end{document}